\documentclass[
reprint,
superscriptaddress,
footinbib,
aps,
amsmath,
amssymb,
pra,
longbibliography
]{revtex4-2}

\usepackage{graphicx} 
\usepackage{mathtools} 
\usepackage{physics} 
\usepackage{bbold}
\usepackage{bm} 
\usepackage{xcolor, soul}
\usepackage{ragged2e}
\usepackage{dsfont}
\usepackage[utf8]{inputenc}
\usepackage[T1]{fontenc}
\usepackage{enumerate}% http://ctan.org/pkg/enumerate

\usepackage{hyperref}
\usepackage{cleveref}

\renewcommand{\Im}{\operatorname{Im}}
\renewcommand{\Re}{\operatorname{Re}}

\crefname{equation}{Eq.}{Eqs.}
\crefname{section}{Sec.}{Secs.}
\crefname{figure}{Fig.}{Figs.}
\crefname{table}{Tab.}{Tabs.}
\crefname{subsection}{Sec.}{Secs.}

\begin{document}
\title{Remote entanglement of massive oscillators via wire-mediated Coulomb interaction}
\author{Lorenzo Papa}
\affiliation{
	Institute for Theoretical Physics and Vienna Center for Quantum Science and Technology, Technical University of Vienna, Wiedner Hauptstraße 8-10, 1040 Vienna, Austria
}
\author{Onur Hosten}
\affiliation{Institute of Science and Technology Austria, Klosterneuburg, Austria}
\author{Carlos Gonzalez-Ballestero}
\email{carlos.gonzalez-ballestero@tuwien.ac.at}
\affiliation{
	Institute for Theoretical Physics and Vienna Center for Quantum Science and Technology, Technical University of Vienna, Wiedner Hauptstraße 8-10, 1040 Vienna, Austria
}
\date{\today}
\begin{abstract}
We propose a method to enhance Coulomb interaction between charged macroscopic mechanical oscillators by placing a conducting structure in their vicinity. We derive the effective motional dynamics of the two oscillators using macroscopic quantum electrodynamics and show that image charges induced in the conductor fundamentally modify the range of the electrostatic interaction. For the specific case of a cylindrical wire, we predict that the coherent motional coupling changes from the free-space scaling $1/D^3$ to an asymptotic $1/(D\ln^2 D)$ dependence on the separation $D$ between the oscillators, at the cost of only negligible additional decoherence for low-frequency oscillators. We further show that, when combined with continuous position measurements, the enhanced interaction enables the generation of steady-state motional entanglement between the oscillators over significantly larger distances than achievable in free space. 
For experimentally realistic milligram-scale oscillators, we predict observable entanglement at separations of several hundred microns -- more than an order of magnitude beyond free-space capabilities -- with improvements approaching two orders of magnitude in future systems. These results identify conductor-assisted Coulomb interactions as a  resource for quantum control of massive objects and for the exploration of entanglement generated by fundamental central forces.
\end{abstract}
\maketitle
\section{Introduction}\label{sec:introduction}
\begin{figure}[t]
	\flushleft
	\includegraphics[width=\linewidth]{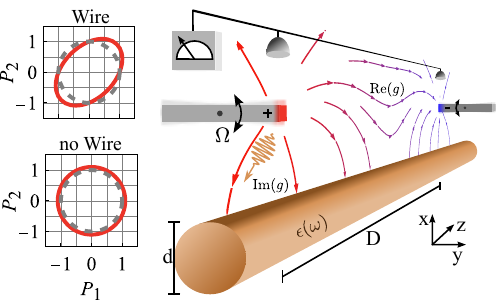}
	\caption{A conducting wire with permittivity $\epsilon(\omega)$ can be used to increase the motional electrostatic coupling between two oscillators, illustrated here as torsional pendula. The image charge distributions generate coherent coupling and dissipation, proportional to real and imaginary parts of the Green's function $g$ respectively. For low-frequency oscillator the latter becomes negligible. Continuous monitoring of the oscillators' positions purifies their steady state, enabling the formation of motional entanglement. Insets: schematic phase-space motional distributions indicating the resulting two-mode squeezing of momentum quadratures $P_{1/2}$.}
	\label{fig:fig1}
\end{figure}
Preparing nonclassical motional states of massive mechanical resonators is a core goal of the optomechanics community~\cite{aspelmeyerCavityOptomechanics2014,bowenQuantumOptomechanics2015}. These states could enable, among others, experiments that probe the quantum nature of gravity~\cite{dewittRoleGravitationPhysics2011}, for instance by the observation of  gravitationally induced entanglement \cite{marlettoGravitationallyInducedEntanglement2017,belenchiaQuantumSuperpositionMassive2018,boseSpinEntanglementWitness2017,christodoulouLocallyMediatedEntanglement2023,martin-martinezWhatGravityMediated2023,bengyatGravitymediatedEntanglementOscillators2024,aspelmeyerQuantumEntanglementGravity2026a}. 
As opposed to entanglement mediated by light~\cite{ockeloen-korppiStabilizedEntanglementMassive2018,riedingerRemoteQuantumEntanglement2018,wollackQuantumStatePreparation2022,kotlerDirectObservationDeterministic2021}, gravitationally induced entanglement experiments are far from feasible with current setups, due to the extremely weak nature of the gravitational coupling \cite{hostenConstraintsProbingQuantum2022,aspelmeyerHowAvoidAppearance2022}, the strong decoherence, and the need to shield the masses from any other interaction \cite{bullingStabilityThresholdsGravitationally2026}. A natural intermediate step toward experiments with gravity is to demonstrate entanglement mediated by the {electrostatic} central force, i.e. the Coulomb interaction~\cite{wuStationaryEntanglementTwo2015,qvarfortMesoscopicEntanglementCentral2020,sohailEnhancedEntanglementInduced2020,coscoEnhancedForceSensitivity2021,rudolphForceGradientSensingEntanglement2022,liEntanglingTwoLevitated2024,deplanoCoulombCouplingTwo2024,winklerSteadystateEntanglementInteracting2025,poddubnyNonequilibriumEntanglementLevitated2025,horovitzParametricResonantEnhancement2026,deyTestingSpontaneousCollapse2026}. However, motional entanglement between massive mechanical oscillators via their mutual Coulomb interaction has not yet been demonstrated. 
One reason is the high level of decoherence experienced by massive oscillators~\cite{frowisMacroscopicQuantumStates2018}. A second reason is that the motional coupling rate between two oscillators decreases with their mass $m$ and with the separation $D$ between their centers of charge as $\propto 1/(mD^3)$.
%A major reason is (i) that the motional coupling rate between two oscillators decreases with their mass $m$ and with the separation $D$ between their centers of charge as $\propto 1/(mD^3)$, as well as (ii) the higher exposure to decoherence \cite{frowisMacroscopicQuantumStates2018}. 
The dependence with $D$ is particularly critical as this separation unavoidably increases with the resonators' size. 
\par     
In this work we propose a route to alleviate this problem by enhancing the range of the Coulomb interaction between two charged oscillators by means of a metallic wire. 
Our approach builds upon related proposals \cite{zurita-sanchezLossyElectricalTransmission2006,daniilidisWiringTrappedIons2009a} developed in trapped ion- and electron quantum science~\cite{anCouplingTwoLaserCooled2022,yuFeasibilityStudyQuantum2022,horneCouplingMotionalQuantum2021,yuStrongCoherentIonelectron2024a}. We describe the wire-oscillator dynamics within the framework of macroscopic electrodynamics \cite{buhmannDispersionForcesMacroscopic2012,martinetzSurfaceInducedDecoherenceHeating2022,jakubecDecoherenceBrownianMotion2025,izadyariSteadyStateEntanglementGeneration2025a} and show that
a high-purity metallic wire can fundamentally modify the scaling of the Coulomb coupling rate, 
while introducing a negligible amount of motional decoherence. 
We then introduce in our model optical continuous measurements of the oscillators' positions \cite{vannerCoolingbymeasurementMechanicalState2013,mikiGeneratingQuantumEntanglement2023a,mikiFeasibleGenerationGravityinduced2024,poddubnyNonequilibriumEntanglementLevitated2025,winklerSteadystateEntanglementInteracting2025}, a known resource that enables to prevent thermalization with the environment and thus reach steady-state entanglement. We predict that recently reported milligram-scale oscillators \cite{agafonovaZigzagOpticalCavity2024a,agafonovaOnemilligramTorsionalPendulum2026} can be entangled at distances of $D\sim$ 0.1 mm, a 10-fold enhancement with respect to free space. We derive analytical expressions for logarithmic negativity and for maximum distance at which entanglement arises, and predict that a 100-fold enhancement in the inter-particle distance $D$ is possible in future systems. Our work not only paves the way toward the observation of quantum effects induced by central forces, but potentially also towards quantum technologies \cite{barzanjehOptomechanicsQuantumTechnologies2022} making use of the quantum control over the Coulomb interaction.\par
This paper is structured as follows. In \cref{sec:system} we derive the master equation describing the dynamics of the two particles in the presence of an arbitrary conducting structure. In \cref{sec:rates} we characterize the coupling and decoherence rates for the specific case of a cylindrical wire. In \cref{sec:impossibleentanglement}, we show on a toy model that modified dynamics are needed to avoid thermalization and to generate stationary entanglement. We therefore introduce a continuous position measurement in \cref{sec:measurementsetup} to purify the steady state and compare the achievable entanglement with and without the wire. Finally, in \cref{sec:conclusion} we summarize our results and discuss future directions.

\section{System and effective dynamics}\label{SectionSystem}

In this section we describe the quantum dynamics of mechanical resonators coupled to a metallic wire. First, in \cref{sec:system}, we derive the effective dynamics of the two resonators after tracing out the wire degrees of freedom. Then, in \cref{sec:rates}, we explicitly calculate all the dynamical rates for a cylindrical wire and characterize the oscillators' dynamics.

\subsection{System description}\label{sec:system}

The system, illustrated in \cref{fig:fig1}, is modeled as two identical point particles with mass $m$ and charge $q$ which are trapped in harmonic potentials with the same frequency $\Omega_0$ and with trap centers placed at $\mathbf{R}_j = (R_j^{x}, 0, R_j^{z})^T$ (j=1,2). The traps have equal x-coordinates $R_1^{x} = R_2^{x}$ and are separated by a distance $D$ along the z-direction ($R_1^{z} = 0$ and $R_2^{z} = D$). Near the particles, an infinitely extended cylindrical wire is placed on the z-axis, with a diameter $d$ and conductivity $\sigma$ (or equivalently relative permittivity ${\epsilon^{\rm wire}_r(\omega) = 1 + \frac{i \sigma}{\omega \epsilon_0}}$, with $\epsilon_0$ the vacuum permittivity). The particles' motion is assumed to be tightly confined along the $y-$ and $z-$ directions, so that one can consider only their motion in the direction perpendicular to the wire. That is, the position operator of both particles is defined as $\hat{\mathbf{R}}_j = (\hat{R}_j^{x}, 0, R_j^{z})$. An extension of the model to three-dimensional motion, arbitrary trap positions, and non-equal mechanical frequencies is straightforward. Note also that our model does also describe spatially extended objects, where $\hat{\mathbf{R}}_j$ would describe the coordinate of the object's center of charge. For the slow and mesoscopic oscillators considered in this work, any changes in the electromagnetic field will  propagate across the system practically instantaneously, i.e. $D \ll (2\pi c)/\Omega_0$, where $c$ is the speed of light. Assuming that there is  no voltage difference applied to the wire ends, the system can then be described in the electroquasistatic limit.\par
Within the electroquasistatic approximation, the Hamiltonian of particles and wire can be decomposed as
\begin{align}\label{equ:Htot}
	\hat{H} = \hat{H}_{\rm p} + \hat{H}_{\rm c} + \hat{H}_{\rm f} + \hat{H}_{\rm pf}.
\end{align}
The first term, $\hat{H}_p = \sum_j (\hat{P}_j^2/(2m) + \frac{1}{2} m \Omega_0^2 {(\hat{R}_j^x - R^x_j)^2})$ is the harmonic Hamiltonian of the trapped particles with $\hat{P}_j$ being the respective particle's momentum operator. The second term ${\hat{H}_{\rm c} = q^2/(4\pi \epsilon_0 \vert\hat{\mathbf{R}}_1-\hat{\mathbf{\mathbf{R}}}_2\vert)}$ denotes the free-space electrostatic Coulomb interaction between the two particles with $\epsilon_0$ denoting the vacuum permittivity. $\hat{H}_{\rm f}$ and $\hat{H}_{\rm pf}$ are the Hamiltonians describing the wire-mediated field and its interaction with the particles, respectively.\par
Following the framework of macroscopic quantum electrodynamics \cite{buhmannDispersionForcesMacroscopic2012, barcellonaManipulatingCoulombInteraction2018} the Hamiltonian of the wire-mediated fields can be written as
\begin{align}\label{equ:field-Hamiltonian}
	\hat{H}_{\rm f} = \int d^3\mathbf{r}\, \int_0^\infty d\omega\,
	\hbar \omega
	\mathbf{\hat{f}}^\dagger(\mathbf{r}, \omega)
	\mathbf{\hat{f}}(\mathbf{r}, \omega),
\end{align}
where $\hbar$ denotes the reduced Planck's constant and
the bosonic fields $\hat{\mathbf{f}}(\mathbf{r},\omega)$ fulfill the commutation relations $[\hat{f}_\alpha(\mathbf{r},\omega),\hat{f}^\dagger_\beta(\mathbf{r'},\omega')] = \delta_{\alpha\beta} \delta(\mathbf{r}-\mathbf{r}') \delta(\omega-\omega')$. The quasi-electrostatic interaction between the particle and the wire-mediated field is then given by
\begin{align}
	\hat{H}_{\rm pf} = \sum_{j=1}^2 q \hat{\Phi}_M(\hat{\mathbf{R}}_j),
\end{align}
where the wire-mediated electric potential $\hat{\Phi}_M$ in the Schr\"odinger picture is defined as
\begin{align}\label{equ:matter-assisted-potential-field}
	\begin{split}
		\hat{\Phi}_M(\mathbf{r})
		 & =
		i \int_0^\infty d\omega \int d^3\mathbf{r}'\,
		\sqrt{
			\frac{\hbar }{\pi\epsilon_0}
			\Im(\epsilon^{\rm wire}_r(\omega))
		}         \\
		 & \times
		\left(\bm{\nabla}' g(\mathbf{r}, \mathbf{r}', \omega)\right) \cdot
		\mathbf{\hat{f}}(\mathbf{r}',\omega) + \rm h.c.
	\end{split}
\end{align}
The medium-assisted electric field can be determined from the above potential as $\hat{\mathbf{E}}_M(\mathbf{r}) = - \bm{\nabla} \hat{\Phi}_M(\mathbf{r})$. The definition \cref{equ:matter-assisted-potential-field} guarantees that both the electric potential and the electric field fulfill their respective fluctuation-dissipation theorems when the fields $\hat{\mathbf{f}}(\mathbf{r},\omega)$ are in a thermal Gibbs state. The complex-valued, frequency-domain Green's function $g(\mathbf{r},\mathbf{r}',\omega)$ solves the electrostatic boundary value problem 
\begin{align}
	\bm{\nabla}\cdot\epsilon_0\epsilon_r(\mathbf{r}, \omega)\bm{\nabla} g(\mathbf{r}, \mathbf{r}', \omega) = - \delta(\mathbf{r}- \mathbf{r}')\, .
	\label{equ:Poisson-Green-equation}
\end{align}
where $\epsilon_r(\mathbf{r},\omega) = \epsilon^{\rm wire}_r(\omega)$ inside the wire and $\epsilon_r(\mathbf{r},\omega) = 1$ in the vacuum outside. The Green's function, especially when both source and field points ($\mathbf{r}$ and $\mathbf{r}'$) are outside the wire, is often split into two parts, $g = g^{\rm fs} + g^{\rm M}$, namely: (i) the free-space Green's function $g^{\rm fs}(\mathbf{r}, \mathbf{r}')$ which already appeared in the free-space Hamiltonian $\hat H_c = q^2 g^{\rm fs}(\hat{\mathbf{R}}_1, \hat{\mathbf{R}}_2)$ and defines the free-space Coulomb electric field as $\mathbf{E}^{\rm fs}(\hat{\mathbf{R}}_1, \hat{\mathbf{R}}_2) = - q \bm{\nabla} g^{\rm fs}(\hat{\mathbf{R}}_1, \hat{\mathbf{R}}_2)$, and (ii) the medium-assisted (also called "scattering") Green's function $g^{\rm M}(\mathbf{r}, \mathbf{r}', \omega)$ which describes the frequency-domain response due to the wire.  \par
Similar to other works \cite{winklerSteadystateEntanglementInteracting2025,martinetzQuantumElectromechanicsLevitated2020}, we first aim at linearizing the position-dependent interaction terms. To this end, note that due to the two interaction terms $\hat H_{\rm c}$ and $\hat H_{\rm pf}$ both the positions of the particles 
$\hat{R}_j^x$ and the wire-mediated fields $\hat{\mathbf{f}}(\mathbf{r},\omega)$ will 
be displaced to new equilibrium values  
$X^{\rm equ}_j$
and $\mathbf{f}^{\rm equ}(\mathbf{r},\omega)$ respectively. We thus expand the Hamiltonian to second order around the unknown 
$X^{\rm equ}_j$ and perform the unitary displacement transformation 
$\hat{\mathbf{f}}(\mathbf{r},\omega) \rightarrow \hat{\mathbf{f}}(\mathbf{r},\omega)+\mathbf{f}^{\rm equ}(\mathbf{r},\omega)$. The expansion is valid provided that 
$
\vert
		E_{x}^{\rm equ}(\mathbf{R}_j^{\rm equ})
	\vert
\ll \langle \hat{X}_j^2 \rangle^{1/2} 
\vert	\partial_{X_k} 
	E_{x}^{\rm equ}(\mathbf{R}_j^{\rm equ})\vert
$ 
where we have defined the deviation from equilibrium  $\hat{X}_j:=\hat{R}_j^x - X^{\rm equ}_{j}$, the equilibrium vector $\mathbf{R}_j^{\rm equ}:=(X_j^{\rm eq},0,R_j^z)^T$, and the classical field
$
\mathbf{E}^{\rm equ} = 
 \mathbf{E}^{\rm fs, equ} + \mathbf{E}^{\rm equ}_M
$. Here $\mathbf{E}^{\rm equ}_M$ denotes the equilibrium value of the medium-assisted electric field, obtained by substituting the operators $\hat{\mathbf{f}}(\mathbf{r},\omega)$ by their equilibrium value $\mathbf{f}^{\rm equ}(\mathbf{r},\omega)$, while similarly $E^{\rm fs, equ} = E^{\rm fs}(\mathbf{R}^{\rm equ}_1, \mathbf{R}^{\rm equ}_2)$.
Finally, the equilibrium values $X^{\rm equ}_{j}$ and fields $\mathbf{f}^{\rm equ}(\mathbf{r},\omega)$ are found by enforcing the zero-force condition, i.e., imposing that the transformed Hamiltonian should not contain linear terms in any operator.
The full derivation is given in Appendix \ref{sec:displacement-trafo}, where we also give the closed nonlinear equations for $X^{\rm equ}_{j}$ and fields $\mathbf{f}^{\rm equ}(\mathbf{r},\omega)$. With this, the Hamiltonian is simplified to
\begin{align}\label{equ:linearized-H}
		\hat{H} \approx \hat{H}_{\rm p}'+ \hat{H}_c^{\rm lin} + \hat{H}_{\rm pf}^{\rm lin} + \hat{H}_{\rm f}
\end{align}
where ${\hat{H}_{\rm p}' = \sum_j (\hat{\mathbf{P}}_j^2/(2m) + m \Omega_1^2 \hat{X}_j^2/2})$  describes the new harmonic oscillator Hamiltonian, including a modified mechanical frequency
\begin{align}
	\begin{split}\label{equ:freq-shift-disp-trafo}
		\Omega_1^2 =  \Omega_0^2 - \frac{q}{m}\bigg(\frac{q}{4 \pi \epsilon_0 D^3}
		              + \partial_{X} E_{M,x}^{\rm equ}(\mathbf{R})
					 \big\vert_{\mathbf{R}=\mathbf{R}^{\rm equ}_j} \bigg).
	\end{split}
\end{align}
This frequency includes a medium-assisted correction, which can be attributed to the interaction of the particles with their image charge distributions \cite{winstoneDirectMeasurementElectrostatic2018}.
The second and third terms in \cref{equ:linearized-H} denote the linearized free-space Coulomb interaction, $\hat{H}_c^{\rm lin} = (q^2/(4 \pi \epsilon_0 D^3)) \hat{X}_1 \hat{X}_2$, and the linearized interaction of the particles with the medium-assisted fields,
$\hat{H}_{\rm pf}^{\rm lin} =  - \sum_{j = 1}^2
	q \hat{X}_j \hat{E}_{\rm M,x}(\mathbf{R}_j^{\rm equ})$. 
\par
Our goal is to derive an effective equation for the two particle dynamics under the influence of the wire as a quantum bath. We thus trace the wire out under a Born-Markov approximation (see Appendix \ref{sec:masterequation-derivation} for details and the end of this section for the validity regime). We assume negligible hybridization of the mechanical modes due to the free-space interaction which is valid when $q^2/(8 \pi \epsilon_0  D^3) \ll m \Omega_1^2$. Furthermore, we assume the low-frequency limit $\hbar \Omega_1 \ll k_{\rm B} T$.
Under these assumptions the master equation takes the following form:
\begin{align}
	\begin{split}
		\frac{d\hat{\rho}}{dt} =
		-\frac{i}{\hbar}\Big[
				\hat{H}_{\rm eff}\, 
				,\hat{\rho}
			\Big]     
		+ \mathcal{D}_{\rm M}[\hat{\rho}]
		+ \mathcal{D}_{\rm th}[\hat{\rho}],
	\end{split}\label{equ:2-particle-Born-Markov-Master-Equation}
\end{align} 
where $\hat{H}_{\rm eff}$ is an effective Hamiltonian, $\mathcal{D}_{\rm M}$ describes the wire-induced decoherence, and we include a thermal dissipator $\mathcal{D}_{\rm th}$ accounting for the intrinsic damping of each particle due to coupling to its own thermal bath. Let us describe each of these terms separately.\par
For convenience, we start by the dissipators. The wire-mediated dissipation is given by
\begin{align}\label{equ:wire-decoherence}
	\mathcal{D}_{\rm M}[\hat{\rho}] = \sum_{j,k=1}^2
	\Gamma_{j k} \Bar{n}
	\Big[
		\hat{x}_j,
		\Big[
			\hat{x}_k,
			\hat{\rho}
			\Big]
		\Big],
\end{align}
with $\Gamma_{jk}\Bar{n}$ appearing as the respective decoherence rates, and with the thermal factor $\Bar{n} = {1/(\exp(\hbar \Omega_1/(k_B T))-1)}$ and the Boltzmann constant $k_{\rm B}$. Here, we have introduced the dimensionless position and momentum quadratures, $\hat{x}_j = \hat{X}_j/x_{\rm zpf}$ and $\hat{p}_j = \hat{P}_j/p_{\rm zpf}$ with the zero-point fluctuations $x_{\rm zpf} = \sqrt{\hbar/(2 m \Omega_1)} = p_{\rm zpf}/(m \Omega_1)$. The diagonal $(j=k)$ terms of \cref{equ:wire-decoherence} are position localization dissipators \cite{schlosshauerQuantumDecoherence2019, gonzalez-ballesteroTheoryCavityCooling2019}, whereas the $j\neq k$ terms generate correlated noise and describe a dissipative coupling between the particles \cite{gonzalez-ballesteroSuppressingRecoilHeating2023b}.
The thermal dissipator in \cref{equ:2-particle-Born-Markov-Master-Equation} takes the the standard finite-temperature damping form given by \cite{breuerTheoryOpenQuantum2002,wallsQuantumOptics2025} 
\begin{equation}\label{equ:thermal-Lindbladian}
	\mathcal{D}_{\rm th}[\hat{\rho}] = {\sum_{j=1}^2 {(\gamma (\Bar{n}+1) \mathcal{L}_{\hat{c}_j\hat{c}_j^\dagger}[\hat{\rho}]} + {\gamma \Bar{n} \mathcal{L}_{\hat{c}_j^\dagger\hat{c}_j}[\hat{\rho}]})},
\end{equation}
where we introduce the mechanical damping rate $\gamma$. In this expression, the ladder operators are defined as $\hat{c}_\alpha=(\hat{x}_\alpha+i\hat{p}_\alpha)/2,$
and the Lindbladian dissipator is given by $\mathcal{L}_{\hat{a} \hat{b}}[\hat{\rho}] = \hat{a}\hat{\rho}\hat{b} - \frac{1}{2}\{\hat{b}\hat{a},\hat{\rho}\}$ with arbitrary operators $\hat{a}$ and $\hat{b}$~\footnote{This dissipator can be derived by tracing out an independent bath of harmonic oscillators, but within the Born-Markov approximation it is justified to simply add it to the master equation as we have done here~\cite{kolodynskiAddingDynamicalGenerators2018}}.\par
The effective Hamiltonian in \cref{equ:2-particle-Born-Markov-Master-Equation} is given by
\begin{align}\label{equ:effective-Hamiltonian}
	\hat{H}_{\rm eff} = 
	\hat{H}_{\rm p}' 
					+ \hat{H}_{\rm c}^{\rm lin} +
					\sum_{j,k=1}^2 \delta\hat{H}_{jk}.
\end{align}
It contains a modification $\delta\hat{H}_{jk}$ due to the presence of the wire, which can be written as
\begin{align}\label{equ:modification-Hamiltonian}
	\delta\hat{H}_{jk}/\hbar =   \mathcal{N}_{jk}\, \hat{x}_j \hat{x}_k + (\Gamma_{jk}/4)(\hat{x}_j\hat{p}_k + \hat{p}_k\hat{x}_j).
\end{align}
Both rates $\mathcal{N}_{jk}$ and $\Gamma_{jk}$ are related to the Green's function through
\begin{align}
	\label{eq:ratesN12}
	\mathcal{N}_{jk} & =
	\frac{q^2 x_{\rm zpf}^2}{2 \hbar}
	\,
	\partial_{X_j} \partial_{X_k} \Re{(g^{\rm M}(\mathbf{R}^{\rm equ}_{j}, \mathbf{R}^{\rm equ}_{k}, \Omega_1))} \\
	\label{eq:ratesGamma}
	\Gamma_{jk}      & = -
	\frac{q^2 x_{\rm zpf}^2}{\hbar}
	\partial_{X_j} \partial_{X_k}
	\Im{(g^{\rm M}(\mathbf{R}^{\rm equ}_{j}, \mathbf{R}^{\rm equ}_{k}, \Omega_1))},
\end{align}
where we have the symmetries $\mathcal{N}_{12} = \mathcal{N}_{21}$, $\mathcal{N}_{11} = \mathcal{N}_{22} = \mathcal{N}$, $\Gamma_{12} = \Gamma_{21}$ and $\Gamma_{11} = \Gamma_{22} = \Gamma$.
The Hamiltonian \cref{equ:modification-Hamiltonian} contains four different wire-induced effects. First, an additional mechanical frequency renormalization $\mathcal{N}$, which rescales the motional frequencies of the particles to the final value
\begin{equation}\label{equ:freq-shift-squeezing-trafo}
    \Omega_f = \sqrt{\Omega_1^2 + 4 \mathcal{N} \Omega_1}.
\end{equation}
Second, a coherent, wire-induced coupling between the motion of the two particles, $\mathcal{N}_{12}$. As we will see below, this new interaction term lies at the heart of our proposal as it will result in longer-range Coulomb interactions. Third, a motional squeezing term $\propto\Gamma$. Fourth, a two-mode-squeezing interaction  $\propto \Gamma_{12}$. 
The total Hamiltonian of \cref{equ:effective-Hamiltonian} can then be written as 
\begin{align}\label{equ:total-Hamiltonian-ladder}
	\begin{split}
		\hat{H}_{\rm eff}/\hbar =
		 & \sum_{j=1}^2 \Omega_f \hat{b}_j^\dagger \hat{b}_j
		+ \frac{G_{\rm tot}}{2}
		(\hat{b}_1+\hat{b}_1^\dagger)(\hat{b}_2+\hat{b}_2^\dagger) \\
		 & + \frac{i}{2}\sum_{j=1}^2 \Gamma_{aa}
		(\hat{b}_j^{\dagger 2}-\hat{b}_j^2)
		+ i\Gamma_{12}(\hat{b}_1^\dagger\hat{b}_2^\dagger-\hat{b}_1\hat{b}_2)
	\end{split}
\end{align}
up to an irrelevant constant. Note that this Hamiltonian is written for convenience in terms of the new ladder operators of the renormalized harmonic potential, $\hat{b}_j=((\hat{x}_j /\sqrt{r})+i\hat{p}_j \sqrt{r})/2$ with
$r = \Omega_1/\Omega_f$. 
The second term in \cref{equ:total-Hamiltonian-ladder} describes the coherent motional coupling at a total rate
\begin{equation}
    G_{\rm tot} = 4r(G_{\rm fs}+\mathcal{N}_{12}),
\end{equation}
which includes the free-space Coulomb coupling $G_{\rm fs} = q^2/(8 \pi \epsilon_0 m \Omega_1 D^3)$ plus a wire-induced contribution $4r\mathcal{N}_{12}$. The single- and two-mode squeezing terms in the second line of \cref{equ:total-Hamiltonian-ladder} can usually be neglected under a rotating wave approximation, which is valid for the parameters considered in this work. 
Notably, all the wire-induced rates in \cref{equ:2-particle-Born-Markov-Master-Equation} are proportional to the ratio $q^2/m$.\par
The master equation \cref{equ:2-particle-Born-Markov-Master-Equation} is applicable to arbitrary geometries of the conductor placed near the particles. It is  valid and all weak-coupling approximations are justified as long as the system is stable and the bath spectral density, i.e. $\partial_{X_j} \partial_{X_k}\Im(g^{\rm M}(\mathbf{R}_j^{\rm equ}, \mathbf{R}_k^{\rm equ}, \omega)$ is linear in $\omega$ in the frequency range spanned by the normal mode frequencies $\Omega_\pm$ defined below (\cref{sec:impossibleentanglement}). Furthermore, the addition of the thermal dissipator \cref{equ:thermal-Lindbladian} is justified as long as $\gamma \ll \Omega_1$.

\subsection{Wire-modified dynamical rates}\label{sec:rates}

Let us evaluate the master equation rates for the cylindrical wire and geometry described in the previous section. We solve the frequency-domain Poisson equation \cref{equ:Poisson-Green-equation} in cylindrical coordinates and find, in agreement with Refs. \cite{jacksonClassicalElectrodynamics2021,cuiElectrostaticPotentialCylindrical2006,eberleinForceNeutralAtom2007} that for the specific case $y=y'=0$, $z-z'=D$ and $x=x'$ the scattering Green's function is given by
\begin{align}\label{eq:Greensfunction}
	\begin{split}
		g^{M}(\mathbf{r}, \mathbf{r}',\omega) =
		-\frac{1}{\pi^2 \epsilon_0}
		\sum\limits_{m=-\infty}^\infty
		\int_0^\infty dk
		\cos(k D) \\
		\times
		\frac{Q_m(k d/2,\omega) }{2}
		\frac{I_m(k d/2)}{K_m(k d/2)}
		K_m(k x)^2,
	\end{split}
\end{align}
where $I_m(\xi)$ and $K_m(\xi)$ are modified Bessel functions of the first and second kind, and where we define the image charge coefficient
% \begin{align}\label{equ:image-charge-coefficient}
% Q_m(\xi, \omega) = \frac{i (\omega \epsilon_0/\sigma)\, A_m(\xi)B_m(\xi) + A_m(\xi)^2}{(\omega^2 \epsilon_0^2/\sigma^2)\,B_m(\xi)^2 + A_m(\xi)^2},
% \end{align}
\begin{align}\label{equ:image-charge-coefficient}
Q_m(\xi, \omega) = \frac{i (\omega \epsilon_0/\sigma)\, B_m(\xi) + 1}{(\omega^2 \epsilon_0^2/\sigma^2)\,B_m(\xi)^2 + 1},
\end{align}
with $B_m(\xi) = 1 - (I_m(\xi) K_m'(\xi))/(I_m'(\xi) K_m(\xi))$.
%with $A_m(\xi) = I_m'(\xi) K_m(\xi)$ and $B_m(\xi) = I_m'(\xi) K_m(\xi) - I_m(\xi) K_m'(\xi)$. 
Note that the imaginary part of the coefficient $Q_m(\xi)$, and with it the imaginary part of the Green's function, is linearly suppressed at low frequencies $\omega \ll \sigma/\epsilon_0$ reflecting the decrease of the Johnson-Nyquist charge fluctuations. As a consequence, for low-frequency oscillators the wire-induced motional decoherence (see Eq.~\eqref{eq:ratesGamma}) becomes negligible. In other words, in this regime the wire behaves as a near-perfect electric conductor that mediates a coherent coupling between the oscillators at the cost of minimal added decoherence. This is a major reason why the wire enables to realize long-range entanglement between the oscillators.
\par
Using the uniform approximation of the modified Bessel functions at large $|m|$, one can show that the integrals will be exponentially suppressed with increasing $|m|$. Thus, to evaluate the Green's function \cref{eq:Greensfunction}, we numerically integrate each pair of terms $\pm m$ in the series starting at $m=0$, and truncate the sum when its relative change is below $10^{-4}$. The upper limit of each k-integral is cutoff at $k_{max} = 25/(2x/d - 1)$, since the integrand is exponentially suppressed for $k > k_{max}$. Special care has to be taken for the $m=0$ integral which is also the dominant part of the sum when $2x/d \gtrsim  1.5$. It has a sharp jump near $k=0$ which becomes especially relevant for large particle separations ($D\gg d$).\par
\begin{figure}[t]
	\centering
	\includegraphics[width=\columnwidth]{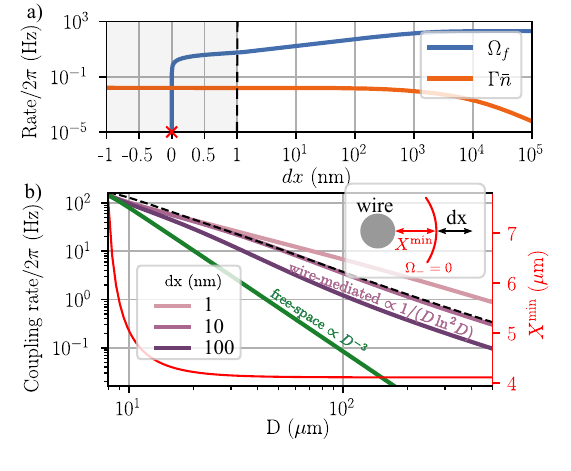}
	\caption{Characterization of the master equation rates for the parameters in \cref{tab:parameters}. (a) Renormalized mechanical frequency $\Omega_f$ and wire-induced decoherence rate $\Gamma \bar{n}$ vs distance to instability $dx$ (see diagram and text for details), (b) Motional particle-particle coupling $G_{\rm tot}$ without the wire (free space, green) and in the presence of the wire (purple) as a function of inter-particle distance $D$. For $dx=10\mathrm{ nm}$, the black dashed line shows the coupling obtained using the analytical asymptotic approximation \cref{eq:asymptoticGreensfunction}. The red solid line shows the minimum equilibrium distance $X^{\rm min}$ as a function of $D$.}. 
	\label{fig:Fig2}
\end{figure}
The behaviour of the master equation rates with respect to inter-particle distance $D$ and radial coordinate $X_1^{\rm equ} = X_2^{\rm equ}$ of the particles is shown in \cref{fig:Fig2} for the parameters listed in \cref{tab:parameters}. Since we are interested in steady-state entanglement, we focus on the regime  $|G_{\rm tot}| \leq \Omega_f$ where the two-particle system is dynamically stable (see \cref{sec:impossibleentanglement}). This is not always the case: specifically, if the original trap centers $\mathbf{R}_j$ are too close to the wire, the wire-induced frequency renormalization can make $\Omega_f$ small enough to destabilize the dynamics. Physically, this represents the Coulomb force between particle and surface charges overcoming the trapping potential, which results in the particle being accelerated towards the wire surface. In other words, stability of the dynamics is only achieved if the equilibrium position is far enough away from the wire. This imposes an effective limitation to the achievable strength of the wire-mediated coupling between particles, and thus to entanglement generation. Within the stable regime we can write the renormalized equilibrium position as $X^{\rm equ}_1=X^{\rm equ}_2 = d/2 +X^{\rm min} + dx$, where $X^{\rm min}$ is the distance between particles and wire surface at which dynamics becomes unstable, and where $dx>0$ (see inset in \cref{fig:Fig2}b). 
The dependence of $X^{\rm min}$ with the distance between the two particles is shown in \cref{fig:Fig2}b. At long distances, when $G_{\rm tot} \ll \Omega_f$, this distance saturates to $X^{\rm min} \approx 4.1\, \mu$m. At short distances it grows sharply due to the strong interaction between the two particles. 
\par
As one decreases $dx$ towards the instability threshold the wire-induced effects grow stronger. These effects are (i) the frequency renormalization of \cref{equ:freq-shift-disp-trafo} and \cref{equ:freq-shift-squeezing-trafo}, (ii) an increase in the wire-mediated decoherence rates $\Gamma_{jk}$ and (iii) an increase in the motional coupling rate $G_{\rm tot}$. To quantify (i) and (ii), we show in \cref{fig:Fig2}(a)  the renormalized mechanical frequency $\Omega_f$ and the decoherence rate $\Gamma \Bar{n}$ as a function of the distance to instability $dx$. Note that, as opposed to $X^{\rm min}$, these two rates do not depend on the separation $D$. The mechanical frequency decreases with $dx$ and eventually becomes $0$ at the instability boundary $dx=0$. In contrast, the decoherence rate $\Gamma \Bar{n}$ saturates to a constant. This enables to work in a near-instability regime incurring only minimal additional decoherence, which, as we will see below, is beneficial for entanglement generation.  
The third effect, namely the modification of the motional coupling rate between the particles $G_{\rm tot}$, is shown in  
\cref{fig:Fig2}(b). Without the wire (green line), the coupling rate shows the characteristic free-space decay $1/D^3$. In the presence of the wire, the coupling becomes orders of magnitude larger at large distances $D$, especially at low radial distances to instability $dx$. As shown in Appendix \ref{sec:asymptoticExpansion}, the scaling of the decay at long distances $D\gg d$ can be derived by asymptotically expanding the Green's tensor in the perfect conductor limit $\Omega_1\ll\sigma/\epsilon_0$. In this limit the wire-induced coherent coupling \cref{eq:ratesN12} becomes
\begin{equation}\label{eq:asymptoticGreensfunction}
    \mathcal{N}_{12}\approx \frac{q^2}{16\pi\epsilon_0 m\Omega_1 (X_1^{\text{ eq}})^2}\frac{1}{D\ln^2(2D/d)},
\end{equation}
resulting in an approximate scaling $G_{\rm tot}\sim 1/(D \ln^2 D)$.
As seen by the dashed line in \cref{fig:Fig2}(b), this approximation accurately recovers the large distance behaviour. 
Note that the coherent coupling rate is largely enhanced by the wire but the wire-induced decoherence remains low. As discussed above, this is a consequence of the low motional frequency of the particles. At such frequencies the wire behaves as a near-perfect electric conductor, able to mediate coherent interactions through the near-lossless motion of the image charges.
In general, the enhancement of the Coulomb interaction at a low decoherence cost is a key feature to attain long-distance entanglement, as we will see below. 

\section{Steady-state entanglement}\label{sec:entanglement}

Having characterized the rates of the master equation, we now assess the extent to which the wire can improve steady-state entanglement generation as compared to the free-space Coulomb interaction. First, in \cref{sec:impossibleentanglement}, we briefly outline why regardless of the coupling strength, steady-state entanglement of low-frequency oscillators enabled only by passive dynamics is impossible at realistic temperatures \cite{ludwigEntanglementMechanicalOscillators2010,qvarfortMesoscopicEntanglementCentral2020,tokarskaGravitationallyInducedEntanglement2025}. Then, in \cref{sec:measurementsetup}, we show how continuous measurement enables to overcome this challenge, and that in combination with the wire, it allows to generate stationary entanglement across much larger distances than in free space.

\subsection{Difficulty of entangling by coupled dynamics}
\label{sec:impossibleentanglement}

First, we want to explore the steady-state of the two-particle system. We start by defining normal mode quadratures as
\begin{align}\label{equ:normal-mode-trafo-a}
	\hat{x}_{\pm} = \sqrt{\Omega_{\pm}/(2\Omega_1)}(\hat{x}_1 \pm \hat{x}_2), \\
	\label{equ:normal-mode-trafo-b}
	\hat{p}_{\pm} = \sqrt{\Omega_1/(2\Omega_{\pm})}(\hat{p}_1 \pm \hat{p}_2),
\end{align} with the two normal mode frequencies ${\Omega_{\pm} = \sqrt{\Omega_f^2\pm G_{\rm tot}\Omega_f}}$. Written in terms of these quadratures, the Hamiltonian \cref{equ:total-Hamiltonian-ladder} splits into two normal mode subspaces, i.e.,  $\hat H_{\rm eff} = \hat H_{\rm eff, +}+\hat H_{\rm eff, -}$ with
$
		\hat{H}_{\rm eff,\pm}/\hbar = 
		\Omega_\pm
		(\hat{x}_\pm^2 + \hat{p}_\pm^2)/4
		+ (\Gamma \pm \Gamma_{12})
		(\hat{x}_\pm\hat{p}_\pm + \rm{h.c.})/4
$. The same is true for all the dissipative terms in \cref{equ:2-particle-Born-Markov-Master-Equation}, so that the dynamics of each of these subspaces is independent. We characterize such dynamics through the quadrature vectors ${\hat{\bm{\eta}}_\pm = (\hat{x}_\pm,  \hat{p}_\pm)^T}$ and the covariance matrices ${\bm\Sigma_{\pm} = \frac{1}{2}\langle \{ \hat{\bm\eta}_\pm, \hat{\bm\eta}_\pm \} \rangle}$ where $\{\mathbf{v}, \mathbf{w}\}_{jk} = (v_j w_k + w_k v_j)$. Their equations of motion are derived from \cref{equ:2-particle-Born-Markov-Master-Equation} and read $\langle \dot{\hat{\bm{\eta}}}_\pm \rangle = A_\pm \langle \hat{\bm{\eta}}_\pm \rangle$ and $\dot{\bm\Sigma}_\pm = A_\pm\bm\Sigma_\pm + \bm\Sigma_\pm A_\pm^T + (2\Bar{n} +1) V_\pm$ respectively, where 
\begin{align}\label{equ:EOMSA}
    A_\pm & =
    \begin{pmatrix}
        -\gamma/2   & \Omega_\pm                          \\
        -\Omega_\pm & -\gamma/2 - \Gamma \pm  \Gamma_{12}
    \end{pmatrix}, \\ \label{equ:EOMSB}
    V_\pm & = \mathrm{diag}(
    \gamma \Omega_\pm/\Omega_1,
    (\gamma + 2 \Gamma \pm 2\Gamma_{12})\Omega_1/\Omega_\pm).
\end{align}
 To derive the above expressions we have also used the relations $\Gamma_{12} = \Gamma_{21}$, as well as $\mathcal{N}_{12} = \mathcal{N}_{21}$ which follow from Onsager reciprocity of the Green's function (see App. \ref{sec:masterequation-derivation}). 
The above equations provide the mathematical condition for the systems' dynamical stability, namely $\max(\text{Re}[\text{eigenvalues}(A_\pm)])<0$. Assuming strong coupling, i.e., $\Gamma_{jk}, \gamma \ll G_{\rm tot}$, this condition reduces to
\begin{align}\label{equ:stability-condition}
	|G_{\rm tot}|<\Omega_f,
\end{align}
which is the condition given in the previous section.
If the system is dynamically stable, a physical steady state exists. Since 
the master equation is quadratic the steady-state is Gaussian and thus fully characterized by the covariance matrix $\bm\Sigma^{\rm ss}_\pm$, as in the steady state $\langle\bm{\hat\eta}^{\rm ss}\rangle=0$. From this covariance matrix a measure of entanglement between the motion of the two particles can be computed. 
A useful entanglement monotone is the logarithmic negativity $\mathcal{E}_N$ which quantifies the violation of the positive-partial-transpose (PPT)-criterion and, for Gaussian states, is positive if and only if the particles are entangled
\cite{vidalComputableMeasureOf2002, alessioserafiniQuantumContinuousVariables2023,duanInseparabilityCriterionContinuous2000,simonPeresHorodeckiSeparabilityCriterion2000}. For our systems's steady state, assuming $G_{\rm tot} \geq 0$ and the ideal regime $\Gamma \Bar{n}, \gamma \Bar{n} \ll G_{\rm tot} \leq \Omega_f$, we find
\begin{align}\label{equ:entanglement}
	\mathcal{E}_N = -\ln\left( \frac{1}{2}(2\Bar{n}+1)
	\sqrt{\frac{4-G_{\rm tot}^2/\Omega_f^2}{G_{\rm tot}/\Omega_f+1}} \right),
\end{align}
which is monotonically increasing with $G_{\rm tot}$ within the stable regime \cref{equ:stability-condition}.
The maximum achievable steady-state logarithmic negativity is thus achieved in the limit $G_{\rm tot}\to\Omega_f$ and is given by $\mathcal{E}_N = - \ln\left(0.612 (2 \Bar{n} + 1)\right)$. Steady-state entanglement $\mathcal{E}_N>0$ then requires a near-zero thermal occupation, specifically $\Bar{n} \leq 0.316$, which is not achievable for most low-frequency oscillators even in cryogenic setups. For instance,  for the parameters in \cref{tab:parameters} steady state entanglement would require a temperature $T \leq 6.7$nK. Crucially, this result is independent of the damping rate which only determines the amount of time it takes to reach the steady state, not which final temperature will be reached. As shown below this limitation can be overcome with the help of continuous position measurements.\par
\begin{table}[ht]
	\centering
	\begin{tabular}{clr}
		\hline
		Parameter       & Description                                                                                                                                 & Value           \\ \hline
		$T$             & temperature                                                                                                                                 & 1 K             \\
		$m$             & mass                                                                                                                                        & 92.5 ng         \\
		$\Omega_0/2\pi$ & bare mechanical frequency                                                                                                                   & 200 Hz          \\
		$\gamma/2\pi$   & bare mechanical damping rate                                                                                                                & $10^{-10}$ Hz   \\
		$q$             & charge    per oscillator                                                                                                                    & $3\times 10^5$e \\
		$\sigma$        & wire conductivity\footnote{corresponds to high-purity copper with RRR = $10^4$ cooled to/under 4.2K \cite{simonPropertiesCopperCopper1992}} & 596 GS/m        \\
		$d$             & wire radius                                                                                                                                 & 5 $\mu$m        \\
	\end{tabular}
	\caption{Parameters used for \cref{fig:Fig2} and \cref{fig:fig3} based on recently demonstrated torsional pendula~\cite{agafonovaZigzagOpticalCavity2024a}, where the effective mass is $m = I/L^2$ with moment of inertia $I$ and length $L$ of the pendulum.}\label{tab:parameters}
\end{table}

\subsection{Entanglement under continuous measurement}\label{sec:measurementsetup}

\begin{figure}[t]
	\centering
	\includegraphics[width=\columnwidth]{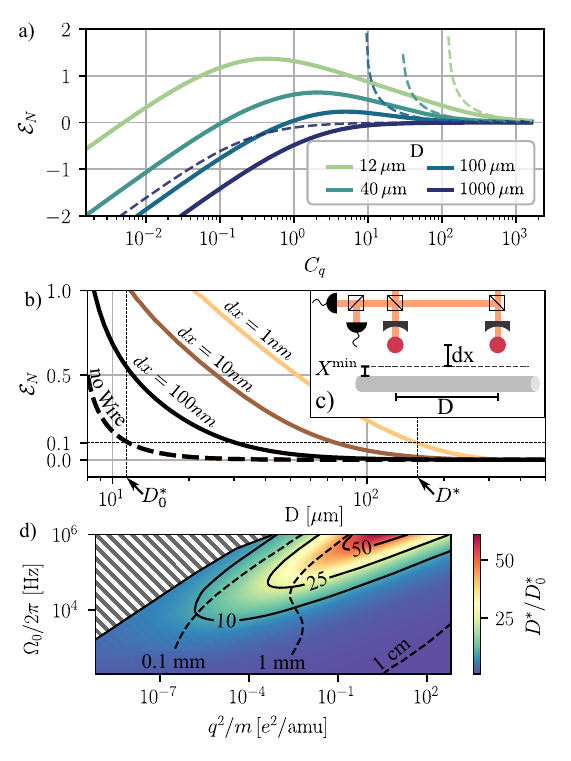}
	\caption{Steady-state entanglement between the two particles under continuous position measurement. (a) Log-negativity $\mathcal{E}_N$ versus quantum cooperativity for different interparticle distances D; dashed lines show the high-cooperativiy analytical approximation \cref{equ:approximateEntanglement}. (b) $\mathcal{E}_N$ versus D at optimal quantum cooperativity and for different radial distances to instability $dx$. (c) Sketch of the proposed setup: the particles act as mirrors of optical cavities and are continuously measured via the cavity output light. (d) Maximum inter-particle distance for which entanglement can be observed, $D^\ast$, normalized by the value in the absence of the wire, $D_0^\ast$, as a function of $q^2/m$ and bare mechanical frequency $\Omega_0$. We choose $dx = \min(10^{-5}X^{\rm min}, 1 \, \rm nm)$ and $\Omega_0/\gamma = 10^{10}$. Solid (dashed) lines show equal value contours of $D^\ast/D^\ast_0$ ($D^\ast$). We exclude the area $X^{\rm min}<0.4\, \mathrm{ \mu m}$ (hatched region) for physical reasons, see main text for details.}
	\label{fig:fig3}
\end{figure}

To achieve stationary entanglement between low-frequency oscillators it is possible to counteract the diffusive trend induced by decoherence by means of measurement. Specifically,
the continuous weak measurement of the oscillators' state with a probe light beam, which causes a continuous update of the state based on the measured outcomes, can purify the conditional mechanical state~\cite{mikiGeneratingQuantumEntanglement2023a,mikiFeasibleGenerationGravityinduced2024,winklerSteadystateEntanglementInteracting2025}. As we will see below, for large enough measurement rates this purification enables the formation of steady-state entanglement. 
To incorporate this measurement in our model, we consider the  system depicted in \cref{fig:fig3}(c), where a probe beam of light is reflected of each oscillator to measure their displacement. Two optical cavities are used to enhance the measurement rate. This configuration is directly implementable with e.g. torsional oscillators \cite{agafonovaZigzagOpticalCavity2024a}, but our results can also be extended to configurations where the oscillator is embedded inside a full optical cavity \cite{purdyObservationRadiationPressure2013,rossiMeasurementbasedQuantumControl2018a,delicMotionalQuantumGround2020,piotrowskiSimultaneousGroundstateCooling2023a,daniaHighpurityQuantumOptomechanics2025}. The two cavities are assumed identical with frequency $\omega_c$, linewidth $\kappa$,  and length $l$. Each cavity is driven resonantly by an external probe laser with power $P_{\rm in}$, and the corresponding input fluctuations around the coherent mean field are assumed to be in vacuum.
This enables to linearize the radiation pressure interaction to  obtain a linear coupling between each particle and its respective cavity mode, at rate  $g = (\omega_c/l)\sqrt{ P_{\rm in}/(2m \Omega_1 \omega_c \kappa)}$ \cite{aspelmeyerCavityOptomechanics2014}. Each cavity is coupled to the outside electromagnetic continuum, which can be detected (for example in a homodyne detection scheme) to perform a weak position measurement on the particles. 
As shown in Appendix \ref{sec:input-output}, in the bad cavity regime $\kappa \gg g$, the cavity mode is uninfluenced by the oscillator and can be adiabatically eliminated from the dynamics. The resulting effective dynamics contains an interaction  between the electromagnetic continuum modes and the position of the oscillators. Specifically, the total Hamiltonian before tracing out the wire degrees of freedom reads 
\begin{equation}\label{eq:measurement-Hamiltonian}
	\hat{H}_{\rm tot}/\hbar = \hat{H}/\hbar + \sum_{j=1}^2 \frac{4g}{\sqrt{\kappa}} \hat{\mathcal{X}}^{\rm in}_j \hat{x}_j,
\end{equation}
where $\hat H$ is given in \cref{equ:linearized-H}, and where $\hat{\mathcal X}^{\rm in}_{j}$, the input-mode quadrature, describes the single collective mode from the continuum that couples to the cavity. This input quadrature, representing the outside field at past times, can be related to the output quadratures -- their long-time analogues denoting the field scattered out of the cavity -- through the input-output relations 
\begin{align}\label{equ:input-output-bad-cavity-quadrature-a}
	&\hat{\mathcal X}^{\rm out}_{j} = \hat{\mathcal X}^{\rm in}_{j},\\
	\label{equ:input-output-bad-cavity-quadrature-b}
	&\hat{\mathcal Y}^{\rm out}_{j} = \hat{\mathcal Y}^{\rm in}_{j} -2g\hat{x}_j,
\end{align}
where $\hat{\mathcal Y}$ denote the phase quadratures of the output field. Note that the output quadratures are the degrees of freedom directly measured in the experiment as they carry the information about the position of the particles $\hat{x}_j$.
Further details can be found in Appendix \ref{sec:input-output}. The procedure detailed in \cref{SectionSystem,sec:impossibleentanglement} can now be repeated using as a starting point the Hamiltonian \cref{eq:measurement-Hamiltonian}. In doing so we obtain the same master equation \cref{equ:2-particle-Born-Markov-Master-Equation}, with a modified Hamiltonian
\begin{align}\label{eq:effective-Hamiltonian-OM}
	\hat{H}_{\rm eff}'/\hbar = \hat H_{\rm eff}/\hbar + \sum_{s={\pm}} \frac{4 g_s}{\sqrt{\kappa}} \hat{x}_s \hat{\mathcal{X}}^{\rm in}_{s},
\end{align}
where $\hat{H}_{\rm eff}$ is given by \cref{equ:total-Hamiltonian-ladder}, and where we have defined the input normal modes as 
$
	\hat{\mathcal{X}}^{\rm in}_{\pm} = (\hat{\mathcal{X}}^{\rm in}_{1} \pm \hat{\mathcal{X}}^{\rm in}_{2})/\sqrt{2}$,
$ 
	\hat{\mathcal{Y}}^{\rm in}_{\pm} = (\hat{\mathcal{Y}}^{\rm in}_{1} \pm \hat{\mathcal{Y}}^{\rm in}_{2})/\sqrt{2}
$ 
and analogously for the output quadratures $\hat{\mathcal{X}}^{\rm out}_{j}$ and $\hat{\mathcal{Y}}^{\rm out}_{j}$. The coupling rate between these quadratures and the mechanical normal modes is given by $g_\pm = g \sqrt{\Omega_1/\Omega_\pm}$.
\par
From this point on, we make use of the theory of quantum filtering \cite{alessioserafiniQuantumContinuousVariables2023, edwardsOptimalQuantumFiltering2005, wisemanQuantumMeasurementControl2009} to model the dynamics of the system under continuous homodyne detection of the output light. When the system of particles and input/output quadratures is in a multimode Gaussian state, the dynamics of the particles under continuous monitoring is described by the classical Kalman-Bucy equations \cite{kalmanNewResultsLinear1961}. A derivation based on \cite{alessioserafiniQuantumContinuousVariables2023} is given in Appendix \ref{sec:ricatti-derivation}. Assuming the two particles are initially uncorrelated, the dynamics of the two normal mode covariance matrices $\bm\Sigma_\pm$ decouple and obey the Riccati equations
\begin{align}
	\begin{split}\label{eq:Riccati}
		\frac{d}{dt}\bm\Sigma_\pm & =  A_\pm\bm\Sigma_\pm + \bm\Sigma_\pm  A_\pm^T + (2\Bar{n}+1) V_\pm \\
		                       & \hspace{1cm} - \bm\Sigma_\pm  C_\pm^T  C_\pm \bm\Sigma_\pm+  H_\pm,
	\end{split}
\end{align}
with $A_\pm$ and $V_\pm$ given by \cref{equ:EOMSA,equ:EOMSB}, while $C_\pm = \mathrm{diag}( 8 g_\pm/\sqrt{\kappa}, 0)$ and $H_\pm = \mathrm{diag}(0, 64 g^2_\pm/\kappa)$. The first line in the above equation describes the evolution in the absence of measurement. The last two terms contain two competing effects induced by the measurement, namely a reduction in the covariances through information gain and a covariance increase due to measurement backaction and input-mode vacuum noise, respectively.
Since we are interested in steady-state entanglement, we set the left-hand-side of \cref{eq:Riccati} to zero and solve the corresponding algebraic Riccati equation. 
From these solutions we compute the logarithmic negativity $\mathcal{E}_N$.
\par
In \cref{fig:fig3}(a) we show the steady-state negativity $\mathcal{E}_N$ as a function of the quantum cooperativity $C_q=g^2/(\kappa \gamma \Bar{n})$, which describes the ratio of measurement rate to decoherence rate, for the parameters of \cref{tab:parameters}. At small interparticle separations $D$ the particles become entangled once the cooperativity is high enough. This is a consequence of the continuous position measurements becoming strong enough to purify the mechanical state to a level where the Coulomb interaction is able to generate entanglement. As cooperativity increases the measurements become stronger and entanglement reaches a maximum at an optimal value of cooperativity. Eventually measurement back-action dominates and suppresses the buildup of correlations, resulting in a decrease of the entanglement. The generated entanglement decreases with distance $D$, which is in line with the decrease in coupling rate shown in \cref{fig:Fig2}(b). For the parameters of \cref{tab:parameters}, at $D \ge 1$ mm the oscillators do not become entangled irrespective of the quantum cooperativity.\par

An analytical approximation for the logarithmic negativity can be found under the assumptions of negligible wire-induced decoherence, $\Gamma_{12},\Gamma \ll \gamma$, high mechanical quality factor, $\Omega_1/\gamma \gg \bar{n} \gg 1$, and high quantum cooperativity $C_q\gg 1$. 
In these limits, we obtain
\begin{align}\label{equ:approximateEntanglement}
	\mathcal{E}_N \approx -\frac{1}{2}\ln\left(
		1 
		+ \frac{1}{32 C_q} 
		- \frac{G_{\rm tot}}{32 \sqrt{2}\, r C_q \gamma \bar{n}}
	\right).
\end{align} 
This expression accurately recovers the exact results at high cooperativities, as shown by the dashed lines in \cref{fig:fig3}(a).
Our analytical approximation also predicts that steady-state entanglement $\mathcal{E}_N > 0$ is only possible under the strong coupling condition $G_{\rm tot} > 4 \sqrt{2}\, \gamma\, k_B T/(\hbar \Omega_f)$. This condition can also be used to estimate the maximum inter-particle distance $D_{\rm max}$ allowing for steady-state entanglement, i.e. the distance at which $\mathcal{E}_N(D_{\rm max})=0$. We do so by introducing the explicit definition of $G_{\rm tot}$ in the asymptotic limit of large separations $D$ derived in Appendix \ref{sec:asymptoticExpansion} and used in \cref{eq:asymptoticGreensfunction}. For $D\gg X^{\rm min}$ the free-space contribution can be neglected and the maximum entangling distance is approximately given by the implicit equation
\begin{align}\label{equ:Dmax}
	D_{\rm max}
	\left(\ln^2\left(\frac{2D_{\rm max}}{d}\right) + \frac{\pi^2}{4}\right)
	 &\approx 
	 \frac{(D_{\rm max}^0)^3 }
	{2 (X^{\rm equ})^2}.
\end{align}
where
\begin{equation}\label{equ:Dmax0}
    D_{\rm max}^0=\sqrt[3]{\frac{q^2}{2\sqrt{2}\pi\epsilon_0m\Omega_1\gamma\bar{n}}}
\end{equation}
is the analogous maximum distance in the absence of the wire.
For the parameters of \cref{fig:fig3}(a) this approximation yields $D_{\rm max} \approx$ 800 $\mu$m which is very close to the exact value.\par
In \cref{fig:fig3}(b) we display the logarithmic negativity at optimal cooperativity as a function of distance $D$ and for different values of $dx$. We observe that the logarithmic negativity is always larger in the presence of the wire than in free space. In addition, the entanglement is larger in magnitude and exists over larger distances $D$ as the distance to the instability $dx$ is decreased, since smaller $dx$ corresponds to a larger coupling $G_{\rm tot}/\Omega_f$. These results are consistent with both the coupling rates shown in \cref{fig:Fig2}(b) and with the approximate expression \cref{equ:approximateEntanglement}. In practice, small amounts of entanglement will not be detectable due to detection noise and other limitations. Therefore, since $\mathcal{E}_N$ in \cref{fig:fig3}(b) is a monotonically decreasing function of the separation $D$, we define the maximum distance at which entanglement can be \emph{observed}, $D^\ast$, as the distance for which the logarithmic negativity is 0.1, that is, $\mathcal E_N(D^\ast)=0.1$. Note that the value $0.1$ is a matter of convention and different detection thresholds can be chosen. The analogue maximum distance for free space, i.e. in the absence of the wire, is labeled $D_0^\ast$. For the parameters of \cref{tab:parameters}
 we obtain $D^\ast/D^\ast_0 = 13.5$ (see the dotted vertical lines in  \cref{fig:fig3}(b)), indicating that the wire enhances the distances at which entanglement can be observed by more than an order of magnitude.\par
To generalize our result to a wider range of parameters, we display in \cref{fig:fig3}(d) the enhancement of the entanglement range, $D^\ast/D^\ast_0$,  as a function of bare mechanical frequency of the oscillators $\Omega_0$ -- i.e., the mechanical frequency far away from the wire -- and of the ratio $q^2/m$ which quantifies the strength of the Coulomb coupling. The cooperativity is fixed to its optimal value and the mechanical quality factor to $Q_0=\Omega_0/\gamma = 10^{10}$. Since fine-tuning the radial distance below a certain resolution is experimentally unfeasible, we fix the distance to instability
as $dx = \min(10^{-5} X^{\rm min}, 1 \, \rm{nm})$. Similarly, since placing the traps too close the wire is experimentally challenging, we discard from the figure the points where $X^{\rm min} + dx< 400$ nm (hashed area).
The remaining parameters ($T$, $\sigma$ and $d$) are taken from \cref{tab:parameters}. 
Generally, for a fixed value of $q^2/m$ the range enhancement factor $D^\ast/D_0^\ast$ increases as a function of $\Omega_0$. We identify the reason as the associated decrease in the thermal occupation and thus in the total decoherence rate $\gamma\bar{n}$. Indeed, we expect that the ratio between observable entanglement distances $D^\ast/D_0^\ast$ behaves in a similar way as the ratio between absolute entanglement distances $D_{\rm max}/D_{\rm max}^0$ defined in \cref{equ:Dmax,equ:Dmax0}, which in the limit of large distance $D_{\rm max}\gg d$ and low decoherence $\gamma \bar{n} \ll q^2/(2\sqrt{2}\pi\epsilon_0m\Omega_1)$ can be shown to increase as  $D_{\rm max}/D_{\rm max}^0\sim (\gamma\Bar{n})^{-2/3}$~\footnote{This can be derived by noting that in this limit, the factor $\pi^2/4$ in \cref{equ:Dmax} can be neglected. This enables to write its analytical solution as $D_{\rm max}=(d/2)\exp[2W\left(\sqrt{(D_{\rm max}^0)^3/d}/(2X_1^{\rm equ})\right)]$ with $W(z)$ the Lambert W function. At low values of decoherence $\gamma\Bar{n}$, the factor $D_{\rm max}^0$
tends to infinity and one can use the asymptotic expansion $W(z)\sim \ln(z)$. This yields, in this limit, $D_{\rm max}\approx (D_{\rm max}^0)^3/(8 (X_1^{\rm{equ }})^2)$ or equivalently $D_{\rm max}\sim 1/(\gamma \Bar{n})$.}. For systems with large values of both $q^2/m$ and $\Omega_0$, $D^\ast/D_0^\ast$ can reach values way above $50$ indicating an enhancement of the entanglement distance of nearly two orders of magnitude. This large enhancement corresponds to parameters typically used in trapped-electron quantum computing \cite{zurita-sanchezWiringSingleElectron2008,yuFeasibilityStudyQuantum2022}.
In general, the results of \cref{fig:fig3}(d) highlight the potential and flexibility of electrical wires to mediate long-range Coulomb interactions between resonators across a vast range of mass- and frequency scales.

\section{Conclusion}\label{sec:conclusion}

We have shown that a conducting wire can substantially extend the range over which steady-state motional entanglement can be generated between charged mechanical oscillators. The enhancement originates from the modified distance dependence of the coherent Coulomb coupling: for a cylindrical wire, the asymptotic scaling changes from the free-space law $\propto 1/D^3$ to a much slower $\propto 1/(D\ln^2D)$ decay. At the same time, for the low-frequency systems considered here, the wire-induced decoherence remains negligible compared to the intrinsic thermal decoherence channels, as the intrinsic resistive losses are negligible at such frequencies. In current milligram-scale optomechanics experiments~\cite{agafonovaZigzagOpticalCavity2024a} this leads to an increase of the entanglement range by a factor of 13.5 relative to free space, a factor that can be largely increased by e.g. increasing the charge of the oscillators or reducing their damping rate. Although our model does not explicitly include decoherence from spurious dielectric surface layers, surface-noise studies in ion traps and charged-particle electromechanics \cite{brownnuttIontrapMeasurementsElectricfield2015,kumphElectricfieldNoiseThin2016,sedlacekDistanceScalingElectricfield2018,martinetzSurfaceInducedDecoherenceHeating2022} suggest that such effects should be negligible at the particle-wire distances considered in this work.
\par
The proof-of-principle enhancements reported here are not fundamental limits, and several routes could further improve them. First, the macroscopic quantum electrodynamics formulation expresses all coherent and dissipative rates in terms of the electrostatic Green's function, providing a natural framework for numerical geometry optimization. For example, sharp conducting features placed near the particles could increase the relevant field gradients, as explored in related ion-based settings \cite{kilianskiAtomsConductingWedge2024a,fountasClassicalQuantumDynamics2019}. Second, parametric driving of the oscillators could be used along with the wire to
amplify the weak motional interactions and thus further increase their entanglement \cite{galveBringingEntanglementHigh2010,poddubnyNonequilibriumEntanglementLevitated2025,shiomatsuBoostingGravityInducedEntanglement2025}. For the specific case of levitated oscillators~\cite{gonzalez-ballesteroLevitodynamicsLevitationControl2021a}, coherent-state expansion \cite{rossiQuantumDelocalizationLevitated2025} can be used in a similar way to enhance transient entanglement~\cite{weissLargeQuantumDelocalization2021}. Finally, note that the conditional entanglement generated by continuous measurements could be complemented by feedback control \cite{mikiGeneratingQuantumEntanglement2023a}. 
This control can be used not only to diagnose unaccounted dynamics such as external forces or colored noise, but also to unconditionally reach the conditional entanglement bound determined in this work. This can be achieved by minimization of the  Einstein-Podolsky-Rosen-type variances~\cite{poddubnyNonequilibriumEntanglementLevitated2025}, possibly in combination with parametric driving. 
\par
On the path toward tabletop tests of gravity-induced entanglement, Coulomb-mediated entanglement offers a controllable setting in which measurement, decoherence, and shielding methods can be studied and tested at much larger coupling strengths. Our work identifies conductor-assisted Coulomb interactions as a flexible resource for exploring such entanglement.
More broadly, the same electrical coupling mechanism is relevant to charged-particle quantum technologies, including levitated electromechanical sensing \cite{goldwaterLevitatedElectromechanicsAllelectrical2019} and wire-coupled trapped ion and electron architectures \cite{zurita-sanchezLossyElectricalTransmission2006,daniilidisWiringTrappedIons2009a,anCouplingTwoLaserCooled2022,horneCouplingMotionalQuantum2021,yuStrongCoherentIonelectron2024a}. 

\begin{acknowledgments}
	 This research was funded in whole or in part by the Austrian Science Fund (FWF) [10.55776/COE1 and PIN3404324]. For Open Access purposes, the author has applied a CC BY public copyright license to any author accepted manuscript version arising from this submission.
     We thank M. Lednev, T.Agrenius and N. Meyer for valuable discussions.
\end{acknowledgments}

\bibliography{bibliography}

@book{alessioserafiniQuantumContinuousVariables2023,
  title = {Quantum {{Continuous Variables}}: {{A Primer}} of {{Theoretical Methods}}},
  shorttitle = {Quantum {{Continuous Variables}}},
  author = {{Alessio Serafini}},
  year = 2023,
  month = aug,
  edition = {2},
  publisher = {CRC Press},
  address = {Boca Raton},
  doi = {10.1201/9781003250975},
  isbn = {978-1-003-25097-5}
}

@article{aspelmeyerCavityOptomechanics2014,
  title = {Cavity Optomechanics},
  author = {Aspelmeyer, Markus and Kippenberg, Tobias J. and Marquardt, Florian},
  year = 2014,
  month = dec,
  journal = {Rev. Mod. Phys.},
  volume = {86},
  number = {4},
  pages = {1391--1452},
  issn = {0034-6861, 1539-0756},
  doi = {10.1103/RevModPhys.86.1391},
  urldate = {2025-06-06},
  copyright = {http://link.aps.org/licenses/aps-default-license},
  langid = {english}
}

@article{barcellonaManipulatingCoulombInteraction2018,
  title = {Manipulating the {{Coulomb}} Interaction: A {{Green}}'s Function Perspective},
  shorttitle = {Manipulating the {{Coulomb}} Interaction},
  author = {Barcellona, Pablo and Bennett, Robert and Buhmann, Stefan Yoshi},
  year = 2018,
  month = mar,
  journal = {J. Phys. Commun.},
  volume = {2},
  number = {3},
  pages = {035027},
  publisher = {IOP Publishing},
  issn = {2399-6528},
  doi = {10.1088/2399-6528/aaa70a},
  urldate = {2025-01-29},
  langid = {english}
}

@article{barzanjehOptomechanicsQuantumTechnologies2022,
  title = {Optomechanics for Quantum Technologies},
  author = {Barzanjeh, Shabir and Xuereb, Andr{\'e} and Gr{\"o}blacher, Simon and Paternostro, Mauro and Regal, Cindy A. and Weig, Eva M.},
  year = 2022,
  month = jan,
  journal = {Nat. Phys.},
  volume = {18},
  number = {1},
  pages = {15--24},
  publisher = {Nature Publishing Group},
  issn = {1745-2481},
  doi = {10.1038/s41567-021-01402-0},
  urldate = {2026-02-06},
  copyright = {2021 Springer Nature Limited},
  langid = {english}
}

@book{bowenQuantumOptomechanics2015,
  title = {Quantum {{Optomechanics}}},
  author = {Bowen, Warwick P. and Milburn, Gerard J.},
  year = 2015,
  month = nov,
  publisher = {CRC Press},
  googlebooks = {YZDwCgAAQBAJ},
  isbn = {978-1-4822-5916-2},
  langid = {english}
}

@book{breuerTheoryOpenQuantum2002,
  title = {The {{Theory}} of {{Open Quantum Systems}}},
  author = {Breuer, Heinz-Peter and Petruccione, Francesco},
  year = 2002,
  publisher = {Oxford University Press},
  googlebooks = {0Yx5VzaMYm8C},
  isbn = {978-0-19-852063-4},
  langid = {english}
}

@book{buhmannDispersionForcesMacroscopic2012,
  title = {Dispersion {{Forces I}}: {{Macroscopic Quantum Electrodynamics}} and {{Ground-State Casimir}}, {{Casimir}}--{{Polder}} and van Der {{Waals Forces}}},
  shorttitle = {Dispersion {{Forces I}}},
  author = {Buhmann, Stefan Yoshi},
  year = 2012,
  series = {Springer {{Tracts}} in {{Modern Physics}}},
  volume = {247},
  publisher = {Springer},
  address = {Berlin, Heidelberg},
  doi = {10.1007/978-3-642-32484-0},
  urldate = {2024-10-01},
  copyright = {https://www.springernature.com/gp/researchers/text-and-data-mining},
  isbn = {978-3-642-32483-3},
  langid = {english}
}

@article{coscoEnhancedForceSensitivity2021,
  title = {Enhanced Force Sensitivity and Entanglement in Periodically Driven Optomechanics},
  author = {Cosco, F. and Pedernales, J. S. and Plenio, M. B.},
  year = 2021,
  month = jun,
  journal = {Phys. Rev. A},
  volume = {103},
  number = {6},
  pages = {L061501},
  publisher = {American Physical Society},
  doi = {10.1103/PhysRevA.103.L061501},
  urldate = {2026-02-09}
}

@article{cuiElectrostaticPotentialCylindrical2006,
  title = {Electrostatic Potential in Cylindrical Dielectric Media Using the Image Charge Method},
  author = {Cui, S. T.},
  year = 2006,
  month = oct,
  journal = {Mol. Phys.},
  volume = {104},
  number = {19},
  pages = {2993--3001},
  publisher = {Taylor \& Francis},
  issn = {0026-8976},
  doi = {10.1080/00268970600926647},
  urldate = {2024-10-17}
}

@article{daniilidisWiringTrappedIons2009a,
  title = {Wiring up Trapped Ions to Study Aspects of Quantum Information},
  author = {Daniilidis, N and Lee, T and Clark, R and Narayanan, S and H{\"a}ffner, H},
  year = 2009,
  month = jul,
  journal = {J. Phys. B: At. Mol. Opt. Phys.},
  volume = {42},
  number = {15},
  pages = {154012},
  issn = {0953-4075},
  doi = {10.1088/0953-4075/42/15/154012},
  urldate = {2026-02-09},
  langid = {english}
}

@article{duanInseparabilityCriterionContinuous2000,
  title = {Inseparability {{Criterion}} for {{Continuous Variable Systems}}},
  author = {Duan, Lu-Ming and Giedke, G. and Cirac, J. I. and Zoller, P.},
  year = 2000,
  month = mar,
  journal = {Phys. Rev. Lett.},
  volume = {84},
  number = {12},
  pages = {2722--2725},
  publisher = {American Physical Society},
  doi = {10.1103/PhysRevLett.84.2722},
  urldate = {2025-01-28}
}

@article{eberleinForceNeutralAtom2007,
  title = {Force on a Neutral Atom near Conducting Microstructures},
  author = {Eberlein, Claudia and Zietal, Robert},
  year = 2007,
  month = mar,
  journal = {Phys. Rev. A},
  volume = {75},
  number = {3},
  pages = {032516},
  publisher = {American Physical Society},
  doi = {10.1103/PhysRevA.75.032516},
  urldate = {2025-02-11}
}

@misc{edwardsOptimalQuantumFiltering2005,
  title = {Optimal {{Quantum Filtering}} and {{Quantum Feedback Control}}},
  author = {Edwards, S. C. and Belavkin, V. P.},
  year = 2005,
  month = aug,
  number = {arXiv:quant-ph/0506018},
  eprint = {quant-ph/0506018},
  publisher = {arXiv},
  doi = {10.48550/arXiv.quant-ph/0506018},
  urldate = {2025-05-09},
  archiveprefix = {arXiv}
}

@article{gonzalez-ballesteroLevitodynamicsLevitationControl2021a,
  title = {Levitodynamics: {{Levitation}} and Control of Microscopic Objects in Vacuum},
  shorttitle = {Levitodynamics},
  author = {{Gonzalez-Ballestero}, C. and Aspelmeyer, M. and Novotny, L. and Quidant, R. and {Romero-Isart}, O.},
  year = 2021,
  month = oct,
  journal = {Science},
  volume = {374},
  number = {6564},
  pages = {eabg3027},
  publisher = {American Association for the Advancement of Science},
  doi = {10.1126/science.abg3027},
  urldate = {2026-07-20}
}

@misc{horneCouplingMotionalQuantum2021,
  title = {Coupling the Motional Quantum States of Spatially Distant Ions Using a Conducting Wire},
  author = {Horne, N. Van and Mukherjee, M.},
  year = 2021,
  month = nov,
  number = {arXiv:2111.14957},
  eprint = {2111.14957},
  primaryclass = {quant-ph},
  publisher = {arXiv},
  doi = {10.48550/arXiv.2111.14957},
  urldate = {2026-02-09},
  archiveprefix = {arXiv}
}

@book{jacksonClassicalElectrodynamics2021,
  title = {Classical {{Electrodynamics}}},
  author = {Jackson, John David},
  year = 2021,
  publisher = {John Wiley \& Sons},
  isbn = {978-1-119-77076-3},
  langid = {english}
}

@article{kalmanNewResultsLinear1961,
  title = {New {{Results}} in {{Linear Filtering}} and {{Prediction Theory}}},
  author = {Kalman, R. E. and Bucy, R. S.},
  year = 1961,
  month = mar,
  journal = {J. Basic Eng},
  volume = {83},
  number = {1},
  pages = {95--108},
  issn = {0021-9223},
  doi = {10.1115/1.3658902},
  urldate = {2026-02-03}
}

@article{kotlerDirectObservationDeterministic2021,
  title = {Direct Observation of Deterministic Macroscopic Entanglement},
  author = {Kotler, Shlomi and Peterson, Gabriel A. and Shojaee, Ezad and Lecocq, Florent and Cicak, Katarina and Kwiatkowski, Alex and Geller, Shawn and Glancy, Scott and Knill, Emanuel and Simmonds, Raymond W. and Aumentado, Jos{\'e} and Teufel, John D.},
  year = 2021,
  month = may,
  journal = {Science},
  volume = {372},
  number = {6542},
  pages = {622--625},
  publisher = {American Association for the Advancement of Science},
  doi = {10.1126/science.abf2998},
  urldate = {2025-02-06}
}

@article{martinetzQuantumElectromechanicsLevitated2020,
  title = {Quantum Electromechanics with Levitated Nanoparticles},
  author = {Martinetz, Lukas and Hornberger, Klaus and Millen, James and Kim, M. S. and Stickler, Benjamin A.},
  year = 2020,
  month = dec,
  journal = {npj Quantum Inf},
  volume = {6},
  number = {1},
  pages = {101},
  publisher = {Nature Publishing Group},
  issn = {2056-6387},
  doi = {10.1038/s41534-020-00333-7},
  urldate = {2026-02-06},
  copyright = {2020 The Author(s)},
  langid = {english}
}

@phdthesis{martinetzQuantumElectromechanicsLevitated2023,
  title = {{Quantum electromechanics with levitated charged particles}},
  author = {Martinetz, Lukas},
  year = 2023,
  month = jun,
  doi = {10.17185/DUEPUBLICO/78558},
  urldate = {2024-10-03},
  collaborator = {{DuEPublico: Duisburg-Essen Publications Online}, University Of Duisburg-Essen and Hornberger, Klaus},
  copyright = {All rights reserved},
  langid = {ngerman},
  school = {DuEPublico: Duisburg-Essen Publications online, University of Duisburg-Essen, Germany}
}

@article{martinetzSurfaceInducedDecoherenceHeating2022,
  title = {Surface-{{Induced Decoherence}} and {{Heating}} of {{Charged Particles}}},
  author = {Martinetz, Lukas and Hornberger, Klaus and Stickler, Benjamin A.},
  year = 2022,
  month = aug,
  journal = {PRX Quantum},
  volume = {3},
  number = {3},
  pages = {030327},
  issn = {2691-3399},
  doi = {10.1103/PRXQuantum.3.030327},
  urldate = {2025-12-04},
  langid = {english}
}

@misc{mikiFeasibleGenerationGravityinduced2024,
  title = {Feasible Generation of Gravity-Induced Entanglement by Using Optomechanical Systems},
  author = {Miki, Daisuke and Matsumura, Akira and Yamamoto, Kazuhiro},
  year = 2024,
  month = may,
  number = {arXiv:2406.04361},
  eprint = {2406.04361},
  primaryclass = {quant-ph},
  publisher = {arXiv},
  doi = {10.48550/arXiv.2406.04361},
  urldate = {2025-08-14},
  archiveprefix = {arXiv}
}

@article{mikiGeneratingQuantumEntanglement2023a,
  title = {Generating Quantum Entanglement between Macroscopic Objects with Continuous Measurement and Feedback Control},
  author = {Miki, Daisuke and Matsumoto, Nobuyuki and Matsumura, Akira and Shichijo, Tomoya and Sugiyama, Yuuki and Yamamoto, Kazuhiro and Yamamoto, Naoki},
  year = 2023,
  month = mar,
  journal = {Phys. Rev. A},
  volume = {107},
  number = {3},
  pages = {032410},
  publisher = {American Physical Society},
  doi = {10.1103/PhysRevA.107.032410},
  urldate = {2026-02-09}
}

@article{ockeloen-korppiStabilizedEntanglementMassive2018,
  title = {Stabilized Entanglement of Massive Mechanical Oscillators},
  author = {{Ockeloen-Korppi}, C. F. and Damsk{\"a}gg, E. and Pirkkalainen, J.-M. and Asjad, M. and Clerk, A. A. and Massel, F. and Woolley, M. J. and Sillanp{\"a}{\"a}, M. A.},
  year = 2018,
  month = apr,
  journal = {Nature},
  volume = {556},
  number = {7702},
  pages = {478--482},
  publisher = {Nature Publishing Group},
  issn = {1476-4687},
  doi = {10.1038/s41586-018-0038-x},
  urldate = {2026-02-09},
  copyright = {2018 Macmillan Publishers Ltd., part of Springer Nature},
  langid = {english}
}

@misc{poddubnyNonequilibriumEntanglementLevitated2025,
  title = {Nonequilibrium Entanglement between Levitated Masses under Optimal Control},
  author = {Poddubny, Alexander N. and Winkler, Klemens and Stickler, Benjamin A. and Deli{\'c}, Uro{\v s} and Aspelmeyer, Markus and Zasedatelev, Anton V.},
  year = 2025,
  month = apr,
  number = {arXiv:2408.06251},
  eprint = {2408.06251},
  primaryclass = {quant-ph},
  publisher = {arXiv},
  doi = {10.48550/arXiv.2408.06251},
  urldate = {2025-05-08},
  archiveprefix = {arXiv}
}

@article{qvarfortMesoscopicEntanglementCentral2020,
  title = {Mesoscopic Entanglement through Central--Potential Interactions},
  author = {Qvarfort, Sofia and Bose, Sougato and Serafini, Alessio},
  year = 2020,
  month = nov,
  journal = {J. Phys. B: At. Mol. Opt. Phys.},
  volume = {53},
  number = {23},
  pages = {235501},
  publisher = {IOP Publishing},
  issn = {0953-4075},
  doi = {10.1088/1361-6455/abbe8d},
  urldate = {2026-02-06},
  langid = {english}
}

@article{riedingerRemoteQuantumEntanglement2018,
  title = {Remote Quantum Entanglement between Two Micromechanical Oscillators},
  author = {Riedinger, Ralf and Wallucks, Andreas and Marinkovi{\'c}, Igor and L{\"o}schnauer, Clemens and Aspelmeyer, Markus and Hong, Sungkun and Gr{\"o}blacher, Simon},
  year = 2018,
  month = apr,
  journal = {Nature},
  volume = {556},
  number = {7702},
  pages = {473--477},
  publisher = {Nature Publishing Group},
  issn = {1476-4687},
  doi = {10.1038/s41586-018-0036-z},
  urldate = {2026-02-09},
  copyright = {2018 Macmillan Publishers Ltd., part of Springer Nature},
  langid = {english}
}

@article{rudolphForceGradientSensingEntanglement2022,
  title = {Force-{{Gradient Sensing}} and {{Entanglement}} via {{Feedback Cooling}} of {{Interacting Nanoparticles}}},
  author = {Rudolph, Henning and Deli{\'c}, Uro{\v s} and Aspelmeyer, Markus and Hornberger, Klaus and Stickler, Benjamin A.},
  year = 2022,
  month = nov,
  journal = {Phys. Rev. Lett.},
  volume = {129},
  number = {19},
  eprint = {2204.13684},
  primaryclass = {quant-ph},
  pages = {193602},
  issn = {0031-9007, 1079-7114},
  doi = {10.1103/PhysRevLett.129.193602},
  urldate = {2025-03-30},
  archiveprefix = {arXiv}
}

@article{simonPeresHorodeckiSeparabilityCriterion2000,
  title = {Peres-{{Horodecki Separability Criterion}} for {{Continuous Variable Systems}}},
  author = {Simon, R.},
  year = 2000,
  month = mar,
  journal = {Phys. Rev. Lett.},
  volume = {84},
  number = {12},
  pages = {2726--2729},
  issn = {0031-9007, 1079-7114},
  doi = {10.1103/PhysRevLett.84.2726},
  urldate = {2025-04-01},
  copyright = {http://link.aps.org/licenses/aps-default-license},
  langid = {english}
}

@misc{winklerSteadystateEntanglementInteracting2025,
  title = {Steady-State Entanglement of Interacting Masses in Free Space through Optimal Feedback Control},
  author = {Winkler, Klemens and Zasedatelev, Anton V. and Stickler, Benjamin A. and Deli{\'c}, Uro{\v s} and {Deutschmann-Olek}, Andreas and Aspelmeyer, Markus},
  year = 2025,
  month = apr,
  number = {arXiv:2408.07492},
  eprint = {2408.07492},
  primaryclass = {quant-ph},
  publisher = {arXiv},
  doi = {10.48550/arXiv.2408.07492},
  urldate = {2025-05-08},
  archiveprefix = {arXiv}
}

@book{wisemanQuantumMeasurementControl2009,
  title = {Quantum {{Measurement}} and {{Control}}},
  author = {Wiseman, Howard M. and Milburn, Gerard J.},
  year = 2009,
  publisher = {Cambridge University Press},
  address = {Cambridge},
  doi = {10.1017/CBO9780511813948},
  urldate = {2025-09-08},
  isbn = {978-0-521-80442-4}
}

@article{yuStrongCoherentIonelectron2024a,
  title = {Strong Coherent Ion-Electron Coupling Using a Wire Data Bus},
  author = {Yu, Baiyi and Betzholz, Ralf and Cai, Jianming},
  year = 2024,
  month = aug,
  journal = {Phys. Rev. Appl.},
  volume = {22},
  number = {2},
  pages = {024032},
  publisher = {American Physical Society},
  doi = {10.1103/PhysRevApplied.22.024032},
  urldate = {2026-02-09}
}

@article{zurita-sanchezLossyElectricalTransmission2006,
  title = {Lossy Electrical Transmission Lines: {{Thermal}} Fluctuations and Quantization},
  shorttitle = {Lossy Electrical Transmission Lines},
  author = {{Zurita-S{\'a}nchez}, Jorge R. and Henkel, Carsten},
  year = 2006,
  month = jun,
  journal = {Physical Review A},
  volume = {73},
  number = {6},
  pages = {063825},
  issn = {1050-2947, 1094-1622},
  doi = {10.1103/PhysRevA.73.063825},
  urldate = {2026-06-19},
  copyright = {http://link.aps.org/licenses/aps-default-license},
  langid = {english}
}

@article{zurita-sanchezWiringSingleElectron2008,
  title = {Wiring up Single Electron Traps to Perform Quantum Gates},
  author = {{Zurita-S{\'a}nchez}, Jorge R. and Henkel, Carsten},
  year = 2008,
  journal = {New Journal of Physics},
  volume = {10},
  number = {8},
  pages = {083021},
  publisher = {IOP Publishing},
  urldate = {2024-08-21}
}

@article{agafonovaOnemilligramTorsionalPendulum2026,
  title = {One-Milligram Torsional Pendulum toward Experiments at the Quantum-Gravity Interface},
  author = {Agafonova, Sofia and Rossell{\'o}, Pere and Mekonnen, Manuel and Hosten, Onur},
  year = 2026,
  month = jan,
  journal = {Communications Physics},
  volume = {9},
  number = {1},
  pages = {80},
  publisher = {Nature Publishing Group},
  issn = {2399-3650},
  doi = {10.1038/s42005-026-02514-w},
  urldate = {2026-06-15},
  copyright = {2026 The Author(s)},
  langid = {english}
}

@article{boseSpinEntanglementWitness2017,
  title = {Spin {{Entanglement Witness}} for {{Quantum Gravity}}},
  author = {Bose, Sougato and Mazumdar, Anupam and Morley, Gavin W. and Ulbricht, Hendrik and Toro{\v s}, Marko and Paternostro, Mauro and Geraci, Andrew A. and Barker, Peter F. and Kim, M. S. and Milburn, Gerard},
  year = 2017,
  month = dec,
  journal = {Physical Review Letters},
  volume = {119},
  number = {24},
  pages = {240401},
  publisher = {American Physical Society},
  doi = {10.1103/PhysRevLett.119.240401},
  urldate = {2026-02-09}
}

@article{vidalComputableMeasureOf2002,
  title = {Computable Measure of Entanglement},
  author = {Vidal, G. and Werner, R. F.},
  year = 2002,
  month = feb,
  journal = {Physical Review A},
  volume = {65},
  number = {3},
  pages = {032314},
  publisher = {American Physical Society},
  doi = {10.1103/PhysRevA.65.032314},
  urldate = {2026-02-16}
}

@article{gonzalez-ballesteroSuppressingRecoilHeating2023b,
  title = {Suppressing {{Recoil Heating}} in {{Levitated Optomechanics Using Squeezed Light}}},
  author = {{Gonzalez-Ballestero}, C. and Zieli{\'n}ska, J.A. and Rossi, M. and Militaru, A. and Frimmer, M. and Novotny, L. and Maurer, P. and {Romero-Isart}, O.},
  year = 2023,
  month = sep,
  journal = {PRX Quantum},
  volume = {4},
  number = {3},
  pages = {030331},
  issn = {2691-3399},
  doi = {10.1103/PRXQuantum.4.030331},
  urldate = {2026-03-11},
  langid = {english}
}

@article{schlosshauerQuantumDecoherence2019,
  title = {Quantum Decoherence},
  author = {Schlosshauer, Maximilian},
  year = 2019,
  month = oct,
  journal = {Physics Reports},
  volume = {831},
  pages = {1--57},
  issn = {03701573},
  doi = {10.1016/j.physrep.2019.10.001},
  urldate = {2026-03-11},
  langid = {english}
}

@article{gonzalez-ballesteroTheoryCavityCooling2019,
  title = {Theory for Cavity Cooling of Levitated Nanoparticles via Coherent Scattering: {{Master}} Equation Approach},
  shorttitle = {Theory for Cavity Cooling of Levitated Nanoparticles via Coherent Scattering},
  author = {{Gonzalez-Ballestero}, C. and Maurer, P. and Windey, D. and Novotny, L. and Reimann, R. and {Romero-Isart}, O.},
  year = 2019,
  month = jul,
  journal = {Physical Review A},
  volume = {100},
  number = {1},
  pages = {013805},
  publisher = {American Physical Society},
  doi = {10.1103/PhysRevA.100.013805},
  urldate = {2025-09-12}
}

@techreport{simonPropertiesCopperCopper1992,
  title = {Properties of Copper and Copper Alloys at Cryogenic Temperatures},
  author = {Simon, N J and Drexler, E S and Reed, R P},
  year = 1992,
  edition = {0},
  number = {NIST MONO 177},
  pages = {NIST MONO 177},
  address = {Gaithersburg, MD},
  institution = {{National Institute of Standards and                                         Technology}},
  doi = {10.6028/NIST.MONO.177},
  urldate = {2025-03-03},
  langid = {english}
}

@article{kolodynskiAddingDynamicalGenerators2018,
  title = {Adding Dynamical Generators in Quantum Master Equations},
  author = {Ko{\l}ody{\'n}ski, Jan and Brask, Jonatan Bohr and {Perarnau-Llobet}, Mart{\'i} and Bylicka, Bogna},
  year = 2018,
  month = jun,
  journal = {Physical Review A},
  volume = {97},
  number = {6},
  pages = {062124},
  issn = {2469-9926, 2469-9934},
  doi = {10.1103/PhysRevA.97.062124},
  urldate = {2026-03-12},
  langid = {english}
}

@misc{NIST:DLMF,
         key = "{\relax DLMF}",
       title = "{\it NIST Digital Library of Mathematical Functions}",
howpublished = "\url{https://dlmf.nist.gov/}, Release 1.2.6 of 2026-03-15",
         url = "https://dlmf.nist.gov/",
        note = "F.~W.~J. Olver, A.~B. {Olde Daalhuis}, D.~W. Lozier, B.~I. Schneider,
                R.~F. Boisvert, C.~W. Clark, B.~R. Miller, B.~V. Saunders,
                H.~S. Cohl, and M.~A. McClain, eds."}

@incollection{kupinVersionWatsonLemma2021,
  title = {A Version of {{Watson}} Lemma for {{Laplace}} Integrals and Some Applications},
  booktitle = {{{EMS Series}} of {{Congress Reports}}},
  author = {Kupin, Stanislas and Naboko, Sergey},
  editor = {Exner, Pavel and Frank, Rupert L. and Gesztesy, Fritz and Holden, Helge and Weidl, Timo},
  year = 2021,
  month = jun,
  edition = {1},
  volume = {18},
  pages = {289--300},
  publisher = {EMS Press},
  doi = {10.4171/ecr/18-1/17},
  urldate = {2026-03-26},
  isbn = {978-3-98547-007-5 978-3-98547-507-0},
  langid = {english}
}

@article{ludwigEntanglementMechanicalOscillators2010,
  title = {Entanglement of Mechanical Oscillators Coupled to a Nonequilibrium Environment},
  author = {Ludwig, Max and Hammerer, K. and Marquardt, Florian},
  year = 2010,
  month = jul,
  journal = {Physical Review A},
  volume = {82},
  number = {1},
  pages = {012333},
  issn = {1050-2947, 1094-1622},
  doi = {10.1103/PhysRevA.82.012333},
  urldate = {2026-04-27},
  copyright = {http://link.aps.org/licenses/aps-default-license},
  langid = {english}
}

@article{frowisMacroscopicQuantumStates2018,
  title = {Macroscopic Quantum States: {{Measures}}, Fragility, and Implementations},
  shorttitle = {Macroscopic Quantum States},
  author = {Fr{\"o}wis, Florian and Sekatski, Pavel and D{\"u}r, Wolfgang and Gisin, Nicolas and Sangouard, Nicolas},
  year = 2018,
  month = may,
  journal = {Reviews of Modern Physics},
  volume = {90},
  number = {2},
  pages = {025004},
  issn = {0034-6861, 1539-0756},
  doi = {10.1103/RevModPhys.90.025004},
  urldate = {2026-04-27},
  langid = {english}
}

@article{yuFeasibilityStudyQuantum2022,
  title = {Feasibility Study of Quantum Computing Using Trapped Electrons},
  author = {Yu, Qian and Alonso, Alberto M. and Caminiti, Jackie and Beck, Kristin M. and Sutherland, R. Tyler and Leibfried, Dietrich and Rodriguez, Kayla J. and Dhital, Madhav and Hemmerling, Boerge and H{\"a}ffner, Hartmut},
  year = 2022,
  month = feb,
  journal = {Physical Review A},
  volume = {105},
  number = {2},
  pages = {022420},
  issn = {2469-9926, 2469-9934},
  doi = {10.1103/PhysRevA.105.022420},
  urldate = {2026-03-25},
  langid = {english}
}

@article{agafonovaZigzagOpticalCavity2024a,
  title = {Zigzag Optical Cavity for Sensing and Controlling Torsional Motion},
  author = {Agafonova, Sofia and Mishra, Umang and Diorico, Fritz and Hosten, Onur},
  year = 2024,
  month = feb,
  journal = {Physical Review Research},
  volume = {6},
  number = {1},
  pages = {013141},
  publisher = {American Physical Society},
  doi = {10.1103/PhysRevResearch.6.013141},
  urldate = {2026-06-15}
}

@article{galveBringingEntanglementHigh2010,
  title = {Bringing {{Entanglement}} to the {{High Temperature Limit}}},
  author = {Galve, Fernando and Pach{\'o}n, Leonardo A. and Zueco, David},
  year = 2010,
  month = oct,
  journal = {Physical Review Letters},
  volume = {105},
  number = {18},
  pages = {180501},
  issn = {0031-9007, 1079-7114},
  doi = {10.1103/PhysRevLett.105.180501},
  urldate = {2026-04-27},
  copyright = {http://link.aps.org/licenses/aps-default-license},
  langid = {english}
}

@article{brownnuttIontrapMeasurementsElectricfield2015,
  title = {Ion-Trap Measurements of Electric-Field Noise near Surfaces},
  author = {Brownnutt, M. and Kumph, M. and Rabl, P. and Blatt, R.},
  year = 2015,
  month = dec,
  journal = {Reviews of Modern Physics},
  volume = {87},
  number = {4},
  pages = {1419--1482},
  publisher = {American Physical Society},
  doi = {10.1103/RevModPhys.87.1419},
  urldate = {2025-01-29}
}

@article{kilianskiAtomsConductingWedge2024a,
  title = {Atoms near a Conducting Wedge: {{Decay}} Rates and Entanglement around a Corner},
  shorttitle = {Atoms near a Conducting Wedge},
  author = {Kilianski, Romuald and Bennett, Robert},
  year = 2024,
  month = dec,
  journal = {Physical Review A},
  volume = {110},
  number = {6},
  pages = {062812},
  publisher = {American Physical Society},
  doi = {10.1103/PhysRevA.110.062812},
  urldate = {2024-12-17}
}

@article{fountasClassicalQuantumDynamics2019,
  title = {Classical and Quantum Dynamics of a Trapped Ion Coupled to a Charged Nanowire},
  author = {Fountas, P. N. and Poggio, M. and Willitsch, S.},
  year = 2019,
  month = jan,
  journal = {New Journal of Physics},
  volume = {21},
  number = {1},
  pages = {013030},
  publisher = {IOP Publishing},
  issn = {1367-2630},
  doi = {10.1088/1367-2630/aaf8f5},
  urldate = {2024-11-27},
  langid = {english}
}

@book{wallsQuantumOptics2025,
  title = {Quantum {{Optics}}},
  author = {Walls, D. F. and Milburn, Gerard J.},
  year = 2025,
  series = {Graduate {{Texts}} in {{Physics}}},
  publisher = {Springer Nature Switzerland},
  address = {Cham},
  doi = {10.1007/978-3-031-84177-4},
  urldate = {2026-04-30},
  copyright = {https://www.springernature.com/gp/researchers/text-and-data-mining},
  isbn = {978-3-031-84176-7 978-3-031-84177-4},
  langid = {english}
}

@book{bechhoeferControlTheoryPhysicists2021,
  title = {Control {{Theory}} for {{Physicists}}},
  author = {Bechhoefer, John},
  year = 2021,
  publisher = {Cambridge University Press},
  address = {Cambridge},
  doi = {10.1017/9780511734809},
  urldate = {2025-09-12},
  isbn = {978-1-107-00118-3}
}

@misc{deyTestingSpontaneousCollapse2026,
  title = {Testing {{Spontaneous Collapse Models}} with {{Coulomb Mediated Squeezing}}},
  author = {Dey, Suroj and Barker, Peter and Datta, Animesh},
  year = 2026,
  month = apr,
  number = {arXiv:2604.21705},
  eprint = {2604.21705},
  primaryclass = {quant-ph},
  publisher = {arXiv},
  doi = {10.48550/arXiv.2604.21705},
  urldate = {2026-05-06},
  archiveprefix = {arXiv}
}

@misc{bullingStabilityThresholdsGravitationally2026,
  title = {Stability {{Thresholds}} for {{Gravitationally Induced Entanglement}} in {{Shielded Setups}}},
  author = {Bulling, Jan and Steiner, Marit O. E. and Pedernales, Julen S. and Plenio, Martin B.},
  year = 2026,
  month = apr,
  number = {arXiv:2604.22593},
  eprint = {2604.22593},
  primaryclass = {quant-ph},
  publisher = {arXiv},
  doi = {10.48550/arXiv.2604.22593},
  urldate = {2026-05-22},
  archiveprefix = {arXiv}
}

@article{wollackQuantumStatePreparation2022,
  title = {Quantum State Preparation, Tomography, and Entanglement of Mechanical Oscillators},
  author = {Wollack, E. Alex and Cleland, Agnetta Y. and Gruenke, Rachel G. and Wang, Zhaoyou and {Arrangoiz-Arriola}, Patricio and {Safavi-Naeini}, Amir H.},
  year = 2022,
  month = apr,
  journal = {Nature},
  volume = {604},
  number = {7906},
  eprint = {2110.07561},
  primaryclass = {quant-ph},
  pages = {463--467},
  issn = {0028-0836, 1476-4687},
  doi = {10.1038/s41586-022-04500-y},
  urldate = {2026-06-15},
  archiveprefix = {arXiv}
}

@article{belenchiaQuantumSuperpositionMassive2018,
  title = {Quantum Superposition of Massive Objects and the Quantization of Gravity},
  author = {Belenchia, Alessio and Wald, Robert M. and Giacomini, Flaminia and {Castro-Ruiz}, Esteban and Brukner, {\v C}aslav and Aspelmeyer, Markus},
  year = 2018,
  month = dec,
  journal = {Physical Review D},
  volume = {98},
  number = {12},
  pages = {126009},
  publisher = {American Physical Society},
  doi = {10.1103/PhysRevD.98.126009},
  urldate = {2026-07-20}
}

@article{christodoulouLocallyMediatedEntanglement2023,
  title = {Locally {{Mediated Entanglement}} in {{Linearized Quantum Gravity}}},
  author = {Christodoulou, Marios and Di Biagio, Andrea and Aspelmeyer, Markus and Brukner, {\v C}aslav and Rovelli, Carlo and Howl, Richard},
  year = 2023,
  month = mar,
  journal = {Physical Review Letters},
  volume = {130},
  number = {10},
  pages = {100202},
  publisher = {American Physical Society},
  doi = {10.1103/PhysRevLett.130.100202},
  urldate = {2026-06-15}
}

@book{dewittRoleGravitationPhysics2011,
  title = {The Role of Gravitation in Physics: Report from the 1957 {{Chapel Hill Conference}}},
  shorttitle = {The Role of Gravitation in Physics},
  author = {DeWitt, C{\'e}cile M. and Rickles, Dean},
  year = 2011,
  publisher = {Edition Open Access},
  urldate = {2026-06-15}
}

@article{martin-martinezWhatGravityMediated2023,
  title = {What Gravity Mediated Entanglement Can Really Tell Us about Quantum Gravity},
  author = {{Mart{\'i}n-Mart{\'i}nez}, Eduardo and Perche, T. Rick},
  year = 2023,
  month = nov,
  journal = {Physical Review D},
  volume = {108},
  number = {10},
  pages = {L101702},
  publisher = {American Physical Society},
  doi = {10.1103/PhysRevD.108.L101702},
  urldate = {2026-06-15}
}

@article{horovitzParametricResonantEnhancement2026,
  title = {Parametric Resonant Enhancement of Motional Entanglement under Optimal Control: {{An}} Analytical Study},
  shorttitle = {Parametric Resonant Enhancement of Motional Entanglement under Optimal Control},
  author = {Horovitz, Gad and Poddubny, Alexander N.},
  year = 2026,
  month = may,
  journal = {Physical Review A},
  volume = {113},
  number = {5},
  pages = {053521},
  publisher = {American Physical Society},
  doi = {10.1103/8lbm-8c9x},
  urldate = {2026-06-15}
}

@article{liEntanglingTwoLevitated2024,
  title = {Entangling Two Levitated Charged Nanospheres through {{Coulomb}} Interaction},
  author = {Li, Guoyao and Yin, Zhangqi},
  year = 2024,
  month = jul,
  journal = {Chinese Physics B},
  volume = {33},
  number = {7},
  pages = {074205},
  publisher = {{Chinese Physical Society and IOP Publishing Ltd}},
  issn = {1674-1056},
  doi = {10.1088/1674-1056/ad3229},
  urldate = {2026-06-15},
  langid = {english}
}

@article{wuStationaryEntanglementTwo2015,
  title = {Stationary Entanglement between Two Nanomechanical Oscillators Induced by {{Coulomb}} Interaction*},
  author = {Wu, Qin and Xiao, Yin and Zhang, Zhi-Ming},
  year = 2015,
  month = nov,
  journal = {Chinese Physics B},
  volume = {25},
  number = {1},
  pages = {014203},
  publisher = {IOP Publishing},
  issn = {1674-1056},
  doi = {10.1088/1674-1056/25/1/014203},
  urldate = {2026-06-15},
  langid = {english}
}

@article{deplanoCoulombCouplingTwo2024,
  title = {Coulomb Coupling between Two Nanospheres Trapped in a Bichromatic Optical Tweezer},
  author = {Deplano, Quentin and Pontin, Antonio and Ranfagni, Andrea and Marino, Francesco and Marin, Francesco},
  year = 2024,
  month = dec,
  journal = {Optica},
  volume = {11},
  number = {12},
  eprint = {2408.02597},
  primaryclass = {physics.optics},
  pages = {1773},
  issn = {2334-2536},
  doi = {10.1364/OPTICA.538760},
  urldate = {2026-06-15},
  archiveprefix = {arXiv}
}

@article{hostenConstraintsProbingQuantum2022,
  title = {Constraints on Probing Quantum Coherence to Infer Gravitational Entanglement},
  author = {Hosten, Onur},
  year = 2022,
  month = jan,
  journal = {Physical Review Research},
  volume = {4},
  number = {1},
  pages = {013023},
  publisher = {American Physical Society},
  doi = {10.1103/PhysRevResearch.4.013023},
  urldate = {2026-06-15}
}

@article{goldwaterLevitatedElectromechanicsAllelectrical2019,
  title = {Levitated Electromechanics: All-Electrical Cooling of Charged Nano- and Micro-Particles},
  shorttitle = {Levitated Electromechanics},
  author = {Goldwater, Daniel and Stickler, Benjamin A and Martinetz, Lukas and Northup, Tracy E and Hornberger, Klaus and Millen, James},
  year = 2019,
  month = jan,
  journal = {Quantum Science and Technology},
  volume = {4},
  number = {2},
  pages = {024003},
  publisher = {IOP Publishing},
  issn = {2058-9565},
  doi = {10.1088/2058-9565/aaf5f3},
  urldate = {2026-06-15},
  langid = {english}
}

@misc{aspelmeyerHowAvoidAppearance2022,
  title = {How to Avoid the Appearance of a Classical World in Gravity Experiments},
  author = {Aspelmeyer, Markus},
  year = 2022,
  month = mar,
  number = {arXiv:2203.05587},
  eprint = {2203.05587},
  primaryclass = {quant-ph},
  publisher = {arXiv},
  doi = {10.48550/arXiv.2203.05587},
  urldate = {2026-06-15},
  archiveprefix = {arXiv}
}

@article{aspelmeyerQuantumEntanglementGravity2026a,
  title = {Quantum Entanglement by Gravity as Tests of Gravitational Collapse Models {\emph{\`a La}} {{Di\'osi}} and {{Penrose}}},
  author = {Aspelmeyer, Markus},
  year = 2026,
  month = jan,
  journal = {Comptes Rendus. Physique},
  volume = {27},
  number = {G1},
  pages = {1--6},
  issn = {1878-1535},
  doi = {10.5802/crphys.270},
  urldate = {2026-06-15},
  langid = {english}
}

@article{bengyatGravitymediatedEntanglementOscillators2024,
  title = {Gravity-Mediated Entanglement between Oscillators as Quantum Superposition of Geometries},
  author = {Bengyat, Ofek and Di Biagio, Andrea and Aspelmeyer, Markus and Christodoulou, Marios},
  year = 2024,
  month = sep,
  journal = {Physical Review D},
  volume = {110},
  number = {5},
  pages = {056046},
  publisher = {American Physical Society},
  doi = {10.1103/PhysRevD.110.056046},
  urldate = {2026-06-15}
}

@article{weissLargeQuantumDelocalization2021,
  title = {Large {{Quantum Delocalization}} of a {{Levitated Nanoparticle Using Optimal Control}}: {{Applications}} for {{Force Sensing}} and {{Entangling}} via {{Weak Forces}}},
  shorttitle = {Large {{Quantum Delocalization}} of a {{Levitated Nanoparticle Using Optimal Control}}},
  author = {Weiss, T. and {Roda-Llordes}, M. and Torrontegui, E. and Aspelmeyer, M. and {Romero-Isart}, O.},
  year = 2021,
  month = jul,
  journal = {Physical Review Letters},
  volume = {127},
  number = {2},
  pages = {023601},
  issn = {0031-9007, 1079-7114},
  doi = {10.1103/PhysRevLett.127.023601},
  urldate = {2026-06-15},
  langid = {english}
}

@article{marlettoGravitationallyInducedEntanglement2017,
  title = {Gravitationally {{Induced Entanglement}} between {{Two Massive Particles}} Is {{Sufficient Evidence}} of {{Quantum Effects}} in {{Gravity}}},
  author = {Marletto, C. and Vedral, V.},
  year = 2017,
  month = dec,
  journal = {Physical Review Letters},
  volume = {119},
  number = {24},
  pages = {240402},
  publisher = {American Physical Society},
  doi = {10.1103/PhysRevLett.119.240402},
  urldate = {2026-06-15}
}

@article{daniaHighpurityQuantumOptomechanics2025,
  title = {High-Purity Quantum Optomechanics at Room Temperature},
  author = {Dania, Lorenzo and Kremer, Oscar Schmitt and Piotrowski, Johannes and Candoli, Davide and Vijayan, Jayadev and {Romero-Isart}, Oriol and {Gonzalez-Ballestero}, Carlos and Novotny, Lukas and Frimmer, Martin},
  year = 2025,
  month = oct,
  journal = {Nature Physics},
  volume = {21},
  number = {10},
  pages = {1603--1608},
  publisher = {Nature Publishing Group},
  issn = {1745-2481},
  doi = {10.1038/s41567-025-02976-9},
  urldate = {2026-06-18},
  copyright = {2025 The Author(s)},
  langid = {english}
}

@article{delicMotionalQuantumGround2020,
  title = {Motional {{Quantum Ground State}} of a {{Levitated Nanoparticle}} from {{Room Temperature}}},
  author = {Deli{\'c}, Uro{\v s} and Reisenbauer, Manuel and Dare, Kahan and Grass, David and Vuleti{\'c}, Vladan and Kiesel, Nikolai and Aspelmeyer, Markus},
  year = 2020,
  month = feb,
  journal = {Science},
  volume = {367},
  number = {6480},
  eprint = {1911.04406},
  primaryclass = {quant-ph},
  pages = {892--895},
  issn = {0036-8075, 1095-9203},
  doi = {10.1126/science.aba3993},
  urldate = {2026-06-18},
  archiveprefix = {arXiv}
}

@article{piotrowskiSimultaneousGroundstateCooling2023a,
  title = {Simultaneous Ground-State Cooling of Two Mechanical Modes of a Levitated Nanoparticle},
  author = {Piotrowski, Johannes and Windey, Dominik and Vijayan, Jayadev and {Gonzalez-Ballestero}, Carlos and Sommer, Andr{\'e}s de los R{\'i}os and Meyer, Nadine and Quidant, Romain and {Romero-Isart}, Oriol and Reimann, Ren{\'e} and Novotny, Lukas},
  year = 2023,
  month = jul,
  journal = {Nature Physics},
  volume = {19},
  number = {7},
  eprint = {2209.15326},
  primaryclass = {quant-ph},
  pages = {1009--1013},
  issn = {1745-2473, 1745-2481},
  doi = {10.1038/s41567-023-01956-1},
  urldate = {2026-06-18},
  archiveprefix = {arXiv}
}

@article{purdyObservationRadiationPressure2013,
  title = {Observation of {{Radiation Pressure Shot Noise}} on a {{Macroscopic Object}}},
  author = {Purdy, T. P. and Peterson, R. W. and Regal, C. A.},
  year = 2013,
  month = feb,
  journal = {Science},
  volume = {339},
  number = {6121},
  eprint = {1209.6334},
  primaryclass = {quant-ph},
  pages = {801--804},
  issn = {0036-8075, 1095-9203},
  doi = {10.1126/science.1231282},
  urldate = {2026-06-18},
  archiveprefix = {arXiv}
}

@article{rossiMeasurementbasedQuantumControl2018a,
  title = {Measurement-Based Quantum Control of Mechanical Motion},
  author = {Rossi, Massimiliano and Mason, David and Chen, Junxin and Tsaturyan, Yeghishe and Schliesser, Albert},
  year = 2018,
  month = nov,
  journal = {Nature},
  volume = {563},
  number = {7729},
  eprint = {1805.05087},
  primaryclass = {quant-ph},
  pages = {53--58},
  issn = {0028-0836, 1476-4687},
  doi = {10.1038/s41586-018-0643-8},
  urldate = {2026-06-18},
  archiveprefix = {arXiv}
}

@article{vannerCoolingbymeasurementMechanicalState2013,
  title = {Cooling-by-Measurement and Mechanical State Tomography via Pulsed Optomechanics},
  author = {Vanner, M. R. and Hofer, J. and Cole, G. D. and Aspelmeyer, M.},
  year = 2013,
  month = aug,
  journal = {Nature Communications},
  volume = {4},
  number = {1},
  pages = {2295},
  publisher = {Nature Publishing Group},
  issn = {2041-1723},
  doi = {10.1038/ncomms3295},
  urldate = {2026-06-18},
  copyright = {2013 Springer Nature Limited},
  langid = {english}
}

@article{anCouplingTwoLaserCooled2022,
  title = {Coupling {{Two Laser-Cooled Ions}} via a {{Room-Temperature Conductor}}},
  author = {An, Da and Alonso, Alberto M. and Matthiesen, Clemens and H{\"a}ffner, Hartmut},
  year = 2022,
  month = feb,
  journal = {Physical Review Letters},
  volume = {128},
  number = {6},
  pages = {063201},
  publisher = {American Physical Society},
  doi = {10.1103/PhysRevLett.128.063201},
  urldate = {2024-09-19}
}

@article{kumphElectricfieldNoiseThin2016,
  title = {Electric-Field Noise above a Thin Dielectric Layer on Metal Electrodes},
  author = {Kumph, Muir and Henkel, Carsten and Rabl, Peter and Brownnutt, Michael and Blatt, Rainer},
  year = 2016,
  month = feb,
  journal = {New Journal of Physics},
  volume = {18},
  number = {2},
  pages = {023020},
  publisher = {IOP Publishing},
  issn = {1367-2630},
  doi = {10.1088/1367-2630/18/2/023020},
  urldate = {2024-11-13},
  langid = {english}
}

@article{sedlacekDistanceScalingElectricfield2018,
  title = {Distance Scaling of Electric-Field Noise in a Surface-Electrode Ion Trap},
  author = {Sedlacek, J. A. and Greene, A. and Stuart, J. and McConnell, R. and Bruzewicz, C. D. and Sage, J. M. and Chiaverini, J.},
  year = 2018,
  month = feb,
  journal = {Physical Review A},
  volume = {97},
  number = {2},
  pages = {020302},
  publisher = {American Physical Society},
  doi = {10.1103/PhysRevA.97.020302},
  urldate = {2025-01-29}
}

@article{sohailEnhancedEntanglementInduced2020,
  title = {Enhanced Entanglement Induced by {{Coulomb}} Interaction in Coupled Optomechanical Systems},
  author = {Sohail, Amjad and Ahmed, Rizwan and Yu, Chang Shui and Munir, Tariq},
  year = 2020,
  month = feb,
  journal = {Physica Scripta},
  volume = {95},
  number = {3},
  pages = {035108},
  publisher = {IOP Publishing},
  issn = {1402-4896},
  doi = {10.1088/1402-4896/ab4dde},
  urldate = {2026-06-15},
  langid = {english}
}

@article{winstoneDirectMeasurementElectrostatic2018,
  title = {Direct Measurement of the Electrostatic Image Force of a Levitated Charged Nanoparticle Close to a Surface},
  author = {Winstone, George and Bennett, Robert and Rademacher, Markus and Rashid, Muddassar and Buhmann, Stefan and Ulbricht, Hendrik},
  year = 2018,
  month = nov,
  journal = {Physical Review A},
  volume = {98},
  number = {5},
  pages = {053831},
  issn = {2469-9926, 2469-9934},
  doi = {10.1103/PhysRevA.98.053831},
  urldate = {2025-01-30},
  langid = {english}
}

@article{izadyariSteadyStateEntanglementGeneration2025a,
  title = {Steady-{{State Entanglement Generation}} via {{Casimir-Polder Interactions}}},
  author = {Izadyari, Mohsen and Pusuluk, Onur and Sinha, Kanu and M{\"u}stecapl{\i}o{\u g}lu, {\"O}zg{\"u}r E.},
  year = 2025,
  month = oct,
  journal = {Scientific Reports},
  volume = {15},
  number = {1},
  eprint = {2406.02270},
  primaryclass = {quant-ph},
  pages = {37105},
  issn = {2045-2322},
  doi = {10.1038/s41598-025-21067-6},
  urldate = {2026-06-19},
  archiveprefix = {arXiv}
}

@misc{tokarskaGravitationallyInducedEntanglement2025,
  title = {Gravitationally {{Induced Entanglement Between Particles}} in {{Harmonic Traps}}: {{Limits}} for {{Gaussian States}}},
  shorttitle = {Gravitationally {{Induced Entanglement Between Particles}} in {{Harmonic Traps}}},
  author = {Tokarska, Julia and Dragan, Andrzej},
  year = 2025,
  month = dec,
  number = {arXiv:2512.24312},
  eprint = {2512.24312},
  primaryclass = {quant-ph},
  publisher = {arXiv},
  doi = {10.48550/arXiv.2512.24312},
  urldate = {2026-01-12},
  archiveprefix = {arXiv}
}

@misc{shiomatsuBoostingGravityInducedEntanglement2025,
  title = {Boosting {{Gravity-Induced Entanglement}} through {{Parametric Resonance}}},
  author = {Shiomatsu, Yuka and Kaku, Youka and Matsumura, Akira and Fujita, Tomohiro},
  year = 2025,
  month = nov,
  number = {arXiv:2511.09169},
  eprint = {2511.09169},
  primaryclass = {gr-qc},
  publisher = {arXiv},
  doi = {10.48550/arXiv.2511.09169},
  urldate = {2025-12-02},
  archiveprefix = {arXiv}
}

@article{jakubecDecoherenceBrownianMotion2025,
  title = {Decoherence and {{Brownian}} Motion of a Polarizable Particle near a Medium},
  author = {Jakubec, Clemens and Jarzynski, Christopher and Sinha, Kanu},
  year = 2025,
  month = oct,
  journal = {Physical Review A},
  volume = {112},
  number = {4},
  pages = {042225},
  publisher = {American Physical Society},
  doi = {10.1103/6m4t-jm4x},
  urldate = {2026-06-19}
}

@article{rossiQuantumDelocalizationLevitated2025,
  title = {Quantum {{Delocalization}} of a {{Levitated Nanoparticle}}},
  author = {Rossi, M. and Militaru, A. and Carlon Zambon, N. and {Riera-Campeny}, A. and {Romero-Isart}, O. and Frimmer, M. and Novotny, L.},
  year = 2025,
  month = aug,
  journal = {Physical Review Letters},
  volume = {135},
  number = {8},
  pages = {083601},
  publisher = {American Physical Society},
  doi = {10.1103/2yzc-fsm3},
  urldate = {2026-06-23}
}

\onecolumngrid

\appendix

\section{Equilibrium shifts and frequency renormalization}\label{sec:displacement-trafo}

The particles' interaction with the electric field shifts the expectation values of both the particle positions $\hat{\mathbf{R}}_j$ and the medium-assisted electric field $\hat{\mathbf{E}}$ and leads to a frequency renormalization from $\Omega_0$ to $\Omega_1$. To find the exact shifts, we first Taylor-expand both $\hat{H}_{\rm c}$ and $\hat{H}_{\rm pf}$ in the full Hamiltonian of \cref{equ:Htot} around $\mathbf{R}_j^{\rm equ}$ and redefine $\hat{R}_j^x := \hat{X}_j + X^{\rm equ}_{j}$. Introducing the displacement between the original and new trap centers,
$
	\Delta_j := R_j^x - X_j^{\rm equ},
$
we obtain
\begin{align}\label{equ:displacement-trafo-starting-point}
	\begin{split}
		\hat{H} = &
		\sum_{j=1}^2 \Bigg(
		\frac{\hat{P}_j^2}{2m} + \frac{1}{2} m \Omega_0^2 \hat{X}_j^2
		+  q \hat{\Phi}_M (\mathbf{R}^{\rm equ}_j)
		- \hat{X}_j
		\left(m \Omega_0^2 \Delta_j +
		\frac{(-1)^{j+1} q^2 (\Delta_2 - \Delta_1) }{4 \pi \epsilon_0 |\mathbf{R}^{\rm equ}_1 - \mathbf{R}^{\rm equ}_2|^3}
		+
		q\hat{E}_{x, \mathrm M}(\mathbf{R}^{\rm equ}_j)
		\right)                                                                  \\
		-         & \hat{X}_j^2 \, \frac{q}{2}
		\left(
		\frac{q}{
			4 \pi \epsilon_0
			|\mathbf{R}_1^{\rm equ} -\mathbf{R}_2^{\rm equ}|^3
		}
		- \frac{3 q(\Delta_2 - \Delta_1)^2}{
			4 \pi \epsilon_0
			|\mathbf{R}_1^{\rm equ} -\mathbf{R}_2^{\rm equ}|^5
		}
		+(\frac{\partial}{\partial x}
		\hat{E}_{x,\mathrm M}(\mathbf{r}))
		\bigg\vert_{
			\mathbf{r}=\mathbf{R}^{\rm equ}_j
		}
		\right)
		\Bigg)                                                                   \\
		+         & \hat{X}_1 \hat{X}_2 \frac{q^2}{4 \pi \epsilon_0}
		\left(
		\frac{1}{
				|\mathbf{R}_1^{\rm equ} -\mathbf{R}_2^{\rm equ}|^3
			}
		-\frac{3 (\Delta_2 - \Delta_1)^2}{
				|\mathbf{R}_1^{\rm equ} -\mathbf{R}_2^{\rm equ}|^5
			}
		\right) + \mathcal{O}(\hat{X}_j^3)
		+ \int d^3\mathbf{r}\int\limits_0^\infty d\omega
		\left(
		\,\hbar \omega\,
		\mathbf{\hat{f}}^\dagger(\mathbf{r}, \omega)
		\cdot\mathbf{\hat{f}}(\mathbf{r}, \omega)
		\right) ,
	\end{split}
\end{align}
where $\hat{\mathbf{f}}$ denotes the bosonic fields introduced under \cref{equ:field-Hamiltonian}, $\hat{\Phi}_M$ and $\hat{E}_M$ are given in (and below) \cref{equ:matter-assisted-potential-field}. Next, we apply a unitary displacement to the bosonic field operators,
\begin{align}
	\hat{\mathbf{f}}(\mathbf{r},\omega) \rightarrow \hat{\mathbf{f}}(\mathbf{r},\omega) + \mathbf{f}^{\rm equ}(\mathbf{r}, \omega).
\end{align}
We then enforce the zero-force condition in \cref{equ:displacement-trafo-starting-point}, which states that all linear terms in $\hat{X}_j$ and $\hat{\mathbf{f}}(\mathbf{r},\omega)$ must vanish. This yields a coupled system of nonlinear equations for $\mathbf{f}^{\rm equ}(\mathbf{r}, \omega)$ and $X_j^{\rm equ}$:
\begin{align}
\begin{split}
	&m \Omega_0^2 \Delta_k +
	\frac{(-1)^{k+1} q^2 (\Delta_2 - \Delta_1) }{4 \pi \epsilon_0 |\mathbf{R}^{\rm equ}_1 - \mathbf{R}^{\rm equ}_2|^3} +\\
	&+
	iq \int_0^\infty d\omega \int d^3\mathbf{r}'\,
		\sqrt{
			\frac{\hbar }{\pi\epsilon_0}
			\Im(\epsilon^{\rm wire}_r(\omega))
		}
		\left( 
			\left(\frac{\partial}{\partial x} g(\mathbf{r}, \mathbf{r}', \omega)\right)\bigg\vert_{\mathbf{r}=\mathbf{R}^{\rm equ}_k} \cdot
			\mathbf{f}^{\rm equ}(\mathbf{r}',\omega) - \rm h.c.
		\right) = 0
\end{split}\\
	&\hbar \omega\, \mathbf{f}^{\rm equ}(\mathbf{r}, \omega) 
	- i q \sum_{k=1}^2 \sqrt{
		\frac{\hbar}{\pi \epsilon_0} 
		\Im(\epsilon^{\rm wire}_r(\omega))
	} 
	\frac{\partial}{\partial x}
	g^*(\mathbf{r}', \mathbf{r}, \omega)
	\bigg\vert_{\mathbf{r}'=\mathbf{R}^{\rm equ}_k} 
	= 0.
\end{align}
Finally, we can make use of the magic formula of the quasi-electrostatic Green's function \cite{martinetzQuantumElectromechanicsLevitated2023}
\begin{align}\label{equ:magic-formula}
	\Im\big(g^{\rm M}(\mathbf{r},\mathbf{r}',\omega)\big)
	= -\epsilon_0 \int d^{3}s\,
	\Im\!\big(\epsilon_r(\omega)\big)
	\bm{\nabla}_s g^{\rm M}(\mathbf{r},\mathbf{s},\omega)
	\cdot
	\bm{\nabla}_s g^{\ast(M)}(\mathbf{s},\mathbf{r}',\omega),
\end{align}
as well as the Kramers-Kronig relations 
\begin{align}\label{equ:Kramers-Kronig-relations}
	\Im(g^{\rm M}(\mathbf{r},\mathbf{r}',\omega)) = 
	-\frac{2}{\pi} \mathcal{P} \int_{0}^\infty d\nu\ \frac{\omega \Re(g^{\rm M}(\mathbf{r},\mathbf{r}',\nu))}{\nu^2 - \omega^2}\\
	\Re(g^{\rm M}(\mathbf{r},\mathbf{r}',\omega)) =
	\frac{2}{\pi} \mathcal{P} \int_{0}^\infty d\nu\ \frac{\nu \Im(g^{\rm M}(\mathbf{r},\mathbf{r}',\nu))}{\nu^2 - \omega^2}
\end{align}
to simplify the system. This yields, on the one hand, the classical electric field 
\begin{align}
	\mathbf{E}^{\rm equ}_{\rm M}(\mathbf{r}) = - q\,\sum_{j=1}^2 \lim_{\omega \rightarrow 0}
	\frac{\partial}{\partial x}  g^{\rm M}(\mathbf{r},\mathbf{R}_j^{\rm equ},\omega),
\end{align} 
that describes the modification of the Coulomb field at $\mathbf{r}$ due to the particle at $\mathbf{r}'$ in the presence of the wire. It is obtained by replacing $\hat{\mathbf{f}}$ with $\mathbf{f}^{\rm equ}$ in the expression for the matter-assisted electric field $\hat{\mathbf{E}}_{\rm M}$ (see \cref{equ:matter-assisted-potential-field}). On the other hand, one finds the classical nonlinear force-balancing equations
\begin{align}\label{equ:force-balance-equation}
	m \Omega_0^2 \Delta_1 + \frac{q^2 (\Delta_2 - \Delta_1)}{4 \pi \epsilon_0 |\mathbf{R}^{\rm equ}_1 - \mathbf{R}^{\rm equ}_2|^3}
	- q \mathbf{E}^{\rm equ}_{\rm M}(\mathbf{R}^{\rm equ}_1) = 0
\end{align}
and the same equation with the labels 1 and 2 swapped. In practice, one finds that $\Delta_1 = \Delta_2$ in the long interparticle-distance limit where $m \Omega_0^2 \gg q^2 \frac{\partial}{\partial x} g(\mathbf{r}, \mathbf{R}_j^{\rm equ}, \omega = 0) \vert_{\mathbf{r} = \mathbf{R}_1^{\rm equ}}>q^2 (\Delta_2 - \Delta_1)/(4 \pi \epsilon_0 D^3)$, because in this limit the equations for the two equilibrium positions decouple and become identical. We numerically solve \cref{equ:force-balance-equation} to obtain the equilibrium positions in all figures of the main text.\par
The frequency renormalization in \cref{equ:freq-shift-disp-trafo} is obtained by similar steps, imposing that all terms in \cref{equ:displacement-trafo-starting-point} proportional to $\hat{X}_j^2$ be written as $\frac{m}{2} \Omega_1^2 \hat{X}_j^2$. Inserting the resulting definitions of $X_j^{\rm equ}$, $\mathbf{f}^{\rm equ}(\mathbf{r},\omega)$ and $\Omega_1$ into the displaced Hamiltonian yields the linearized Hamiltonian in \cref{equ:linearized-H}.

\section{Master equation derivation}\label{sec:masterequation-derivation}

We start by grouping the Hamiltonian after the displacement transformation, see \cref{equ:linearized-H}, into a system part S, bath part B, and interaction part:
\begin{align}\label{equ:linearizedHamiltonian-grouped}
	\hat{H} = \underbrace{\hat{H}_{\rm p}' + \hat{H}^{\rm lin}_c}_{\hat{H}_{\rm S}} + \underbrace{\hat{H}_{\rm f}}_{\hat{H}_{\rm B}} + \underbrace{\hat{H}^{\rm lin}_{\rm pf}}_{\hat{H}_{\rm int}}.
\end{align}
When the coupling between system and bath is weak (which we quantify at the end of \cref{sec:system}), we may undertake the Born-Markov approximation. Specifically, this approximates the state as separable at all times, $\hat \rho = \hat \rho_S \otimes \hat \rho_B^{\rm th}$ with the bath state $\hat \rho_B^{\rm th} = e^{-\hat H_B/(k_B T)}/Z$ remaining in thermal Gibbs state, where $Z$ denotes the partition function. It also approximates the system reduced dynamics as being local in time, under the assumption that the bath correlators decay on a much faster time scale $\tau_B$ than the system-bath coupling. 
The resulting master equation in the interaction picture -- denoted by the super-index ``I'' --reads \cite{breuerTheoryOpenQuantum2002} 
\begin{align}
	\begin{split}
		\frac{d}{dt}\hat \rho_{\rm S}^{I}(t)
		=  - \frac{1}{\hbar^2}\int_0^\infty d\tau\,
		\Tr_{\rm B}\!\Bigg(
		\bigg[\hat H_{\rm int}^{I}(t),
				\comm{\hat H_{\rm int}^{I}(t-\tau)}{\hat \rho_{\rm S}^{I}(t)\otimes \hat \rho_{\rm B}}\bigg]
		\Bigg).
	\end{split}
\end{align}
We proceed by inserting the interaction Hamiltonian $\hat{H}_{\rm pf}^{\rm lin}$ (given below \cref{equ:linearized-H}), which yields 
\begin{equation}\label{equ:redfield-bath-correlators}
    \frac{d}{dt}\hat{\rho}_{\rm S}^{I}(t)
		= -\frac{1}{\hbar^2}\sum_{j,k=1}^2 \int_0^\infty d\tau\,
		\Big(
		 C_{jk}(\tau)
		\left[
			\hat{X}_j^I(t),
			\hat{X}_k^I(t-\tau)\hat{\rho}_{\rm S}^I(t)
		\right] + C_{kj}(-\tau)
		\left[
			\hat{\rho}_{\rm S}^I(t)\hat{X}_k^I(t-\tau),
			\hat{X}_j^I(t)
		\right]
		\Big).
\end{equation}
The bath correlators are defined as
\begin{align}\label{equ:bath-correlators}
	C_{jk}(\tau)
	=
	q^2 \int_0^\infty d\omega \int_0^\infty d\omega' 
	\langle
		(
			\hat{E}_{\rm M,x}(
				\mathbf{R}_j^{\rm equ}, 
				\omega) 
			 + \rm{h.c.}
		)
		(
			\hat{E}_{\rm M,x}(
				\mathbf{R}_k^{\rm equ}, 
				\omega'
			) 
			e^{i \omega' \tau} 
			+ \rm{h.c.}
		)
		\rangle_{\rm th},
\end{align}
where the medium-assisted, Schr\"odinger-picture electric field operator has been split into positive- and negative-frequency components as
\begin{align}
	\hat{E}_{\rm M,x}(\mathbf{r})
	=
	\int_0^\infty d\omega\,
	\left[
		\hat{E}_{\rm M,x}(\mathbf{r},\omega)
		+
		\hat{E}_{\rm M,x}^{\dagger}(\mathbf{r},\omega)
	\right],
\end{align}
and the definition of $\hat{E}_{\rm M,x}(\mathbf{r},\omega)$ can be found under \cref{equ:matter-assisted-potential-field}.
The bath-correlators of \cref{equ:bath-correlators} are evaluated using the fluctuation-dissipation relation of the electric field \cite{buhmannDispersionForcesMacroscopic2012, martinetzQuantumElectromechanicsLevitated2023},
\begin{align}
	\begin{split}
		\langle \hat{\mathbf{E}}^\dagger(\mathbf{r},\omega) \otimes \hat{\mathbf{E}}(\mathbf{r'},\omega') \rangle_{\rm th} =
		- \frac{\hbar}{\pi} \delta(\omega-\omega') \Bar{n}(\omega,T)\bm{\nabla} \otimes \bm{\nabla}' \Im(g^{M}(\mathbf{r},\mathbf{r}', \omega)),
	\end{split}
\end{align}
as well as $\langle \hat{\mathbf{E}}(\mathbf{r},\omega) \otimes \hat{\mathbf{E}}(\mathbf{r'},\omega') \rangle_{\rm th} = 0$. This results in 
\begin{align}\label{equ:evaluated-bath-correlators}
	C_{jk}(\tau)
	=
	-\frac{\hbar q^2}{\pi}
	\int_0^\infty d\omega\,
	\partial_{X_j}\partial_{X_k}
	\Im\!\left[
		g\!\left(
		\mathbf{R}_j^{\rm equ},
		\mathbf{R}_k^{\rm equ},
		\omega
		\right)
	\right]
	\left[
		\left(\Bar{n}(\omega,T)+1\right)e^{-i\omega\tau}
		+
		\Bar{n}(\omega,T)e^{i\omega\tau}
	\right].
\end{align}
The Born-Markov approximation remains valid as long as this correlator decays faster than the system-bath interaction time $\tau_B$. As is well known, this is the case for Ohmic bath spectral densities \cite{breuerTheoryOpenQuantum2002}. In our case, this would correspond to a linear frequency dependence of $\Im\!\left[
		g\!\left(
		\mathbf{R}_j^{\rm equ},
		\mathbf{R}_k^{\rm equ},
		\omega
		\right)
	\right]$.\\
To further simplify the master equation \cref{equ:redfield-bath-correlators}, we note that in the large distance limit that we are primarily intrested in the free-space Coulomb interaction will be very weak, $ q^2/(8 \pi \epsilon_0  D^3) \ll m \Omega_1^2$, so that the interaction picture system operators $\hat{X}_j^I(t)=x_{\rm zpf}(\hat{b}_j(t)+\hat{b}_j^\dagger(t))$ will not hybridize. This allows us to write $e^{\frac{i}{\hbar} \hat{H}_{\rm S} t} \hat{b}_j e^{-\frac{i}{\hbar} \hat{H}_{\rm S} t} \approx e^{-i \Omega_1 t} \hat{b}_j$. The Schrödinger-picture master equation then takes the following form after inserting the evaluated bath correlators of (\cref{equ:evaluated-bath-correlators}),
\begin{align}\label{equ:redfield-expanded-bath-correlators}
	\begin{split}
		&\frac{d}{dt}\hat{\rho}_{\rm S}(t)
		= 
		- \frac{i}{\hbar} \comm{
			\hat{H}_S
		}{
			\hat{\rho}_{\rm S}(t)
		} + 
		\frac{q^2 x_{\rm zpf}^2}{\pi\hbar}
		\sum_{j,k=1}^2
		\int_0^\infty d\omega\,
		\partial_{X_j}\partial_{X_k}
		\Im\!\left[
			g^{\rm M}\!\left(
				\mathbf{R}_j^{\rm equ},
				\mathbf{R}_k^{\rm equ},
				\omega
				\right)
				\right] 
				\int_0^\infty d\tau\\
		&\times
		\Bigg(
        \left( 
            \hat{b}_j 
            \hat{b}_k 
            \hat{\rho}_{\rm S}
            e^{i\Omega_1 \tau}
            + 
            \hat{b}_j 
            \hat{b}^\dagger_k 
            \hat{\rho}_{\rm S}
            e^{-i\Omega_1 \tau}
            + 
            \hat{b}^\dagger_j 
            \hat{b}_k 
            \hat{\rho}_{\rm S}
            e^{i\Omega_1 \tau}
            + 
            \hat{b}^\dagger_j 
            \hat{b}^\dagger_k 
            \hat{\rho}_{\rm S}
            e^{-i\Omega_1 \tau}
        \right)
            \left( 
                e^{-i\omega \tau} 
                (\bar{n}(\omega,T)+1)  
                + 
                e^{+i\omega \tau} 
                \bar{n}(\omega,T) 
            \right)\\
        &- 
        \left( 
            \hat{b}_j 
            \hat{\rho}_{\rm S}
            \hat{b}_k 
            e^{i\Omega_1 \tau}
            + 
            \hat{b}_j 
            \hat{\rho}_{\rm S} 
            \hat{b}^\dagger_k 
            e^{-i\Omega_1 \tau}
            + 
            \hat{b}^\dagger_j 
            \hat{\rho}_{\rm S}
            \hat{b}_k 
            e^{i\Omega_1 \tau}
            + 
            \hat{b}^\dagger_j 
            \hat{\rho}_{\rm S}
            \hat{b}^\dagger_k 
            e^{-i\Omega_1 \tau}
        \right)
            \left( 
                e^{+i\omega \tau} 
                (\bar{n}(\omega,T)+1)  
                + 
                e^{-i\omega \tau} 
                \bar{n}(\omega,T) 
            \right)
    +  H.c. \Bigg).
	\end{split}
\end{align}
We then use the Sokhotski-Plemelj relation
\begin{equation}
	\int_0^\infty d\tau e^{i (\omega \pm \nu) \tau} = \pi \delta(\omega\pm\nu) + i\mathcal{P}\frac{1}{\omega \pm \nu}
\end{equation}
(with $\mathcal{P}$ indicating the Cauchy principal value) to evaluate the $\tau$-integrals, which yields the master equation
\begin{align}\label{equ:master-equation-2particle-v0}
	\begin{split}
		\frac{d}{dt}\rho_{\rm S} = -\frac{i}{\hbar}
		\comm{
			\hat{H}_{\rm p}' + \hat{H}_{\rm c}^{\rm lin} + \sum_{j,k=1}^2 \delta\hat{H}_{jk}
		}{
			\hat{\rho}_{\rm S}
		}
		 & + \sum_{j,k=1}^2 \Bigg(
		-\Gamma_{jk} \left((\bar{n}(\Omega_1,T)+1) \mathcal{D}_{\hat{x}_j \hat{b}^\dagger_k}[\hat{\rho}_{\rm S}] + \bar{n}(\Omega_1,T) \mathcal{D}_{\hat{x}_j \hat{b}_k}[\hat{\rho}_{\rm S}] + \mathrm{h. c.}\right) \\
		 & + i \mathcal{N}_{jk}\left(
		\mathcal{D}_{\hat x_j \hat x_k}[\hat \rho]
		-\mathcal{D}_{\hat x_k \hat x_j}[\hat \rho]
		\right)
		- \mathcal{S}_{jk}
		\left(
		\mathcal{D}_{\hat{x}_j \hat{p}_k}[\hat{\rho}_{\rm S}] +
		\mathcal{D}_{\hat{p}_k \hat{x}_j}[\hat{\rho}_{\rm S}]
		\right)
		\Bigg)
	\end{split}
\end{align}
with the modification to the Hamiltonian $\delta\hat{H}_{jk}$ given in \cref{equ:modification-Hamiltonian}
, rates $\Gamma_{jk}$ defined already in \cref{eq:ratesGamma}, and with
\begin{align}
	\label{equ:rateSij}
	\mathcal{S}_{jk} & = \mathcal{P}\int\limits_0^\infty d\omega\,
	\frac{q^2 x_{\rm zpf}^2}{\hbar} \,
	\frac{
		\Omega_1 (2\Bar{n}(\omega)+1)
	}{
		\omega^2 - \Omega_1^2
	}
	\,\partial_{X_j} \partial_{X_k} \Im(g^{M}(
	\mathbf{R}_j^{\rm equ},
	\mathbf{R}_k^{\rm equ},
	\omega
	))                                                             \\
	\mathcal{N}_{jk} & = \int_0^\infty d\omega\,
	\frac{q^2 x_{\rm zpf}^2}{\pi \hbar}
	\frac{\omega}{\omega^2 - \Omega_1^2}\,
	\,
	\partial_{X_j} \partial_{X_k} \Im(g^{M}(
	\mathbf{R}_j^{\rm equ},
	\mathbf{R}_k^{\rm equ},
	\omega
	)).
\end{align}
The master equation \cref{equ:master-equation-2particle-v0} can be simplified further. First, to simplify the dissipator into \cref{equ:wire-decoherence} we make the high temperature approximation ($\Bar{n}\gg1$) so that $\bar{n} \approx \bar{n} +1$, which leads to
\begin{align*}
	-\Gamma_{jk} \left((\Bar{n}+1) \mathcal{D}_{\hat{x}_j \hat{b}^\dagger_k}[\hat{\rho}_{\rm S}] + \Bar{n} \mathcal{D}_{\hat{x}_j \hat{b}_k}[\hat{\rho}_{\rm S}] + \mathrm{h. c.}\right)
	\mapsto
	\Gamma_{jk} \Bar{n}
	\Big[\hat{x}_j, \Big[ \hat{x}_k, \hat{\rho} \Big]\Big].
\end{align*}
Secondly, since the medium assisted
scattering part $g^{\rm M}$ is a causal response function, it satisfies the Kramers-Kronig relations \cref{equ:Kramers-Kronig-relations}. Using them simplifies the expression for $\mathcal{N}_{jk}$ to the one given in \cref{eq:ratesN12}. Furthermore, the
particles are assumed to have the same charge, mass, and frequency which, together with
Onsager reciprocity of the Green's function \cite{martinetzQuantumElectromechanicsLevitated2023} which reads $\partial_{X_j} \partial_{X_k} g^{\rm M}(
	\mathbf{R}_j^{\rm equ},
	\mathbf{R}_k^{\rm equ},
	\omega
	) = \partial_{X_k} \partial_{X_j} g^{\rm M}(
	\mathbf{R}_k^{\rm equ},
	\mathbf{R}_j^{\rm equ},
	\omega
)$, 
makes all the master equation rates $\Gamma_{jk}$, $\mathcal{N}_{jk}$ and $\mathcal{S}_{jk}$ symmetric under the exchange of $j$ and $k$. This results in 
\begin{align}
i \sum_{jk}\mathcal{N}_{jk}\left(
		\mathcal{D}_{\hat x_j \hat x_k}[\hat \rho]
		-\mathcal{D}_{\hat x_k \hat x_j}[\hat \rho]
		\right) = 0,
\end{align}
as well as 
\begin{align}
\sum_{jk} \mathcal{S}_{jk}
		\left(
		\mathcal{D}_{\hat{x}_j \hat{p}_k}[\hat{\rho}_{\rm S}] +
		\mathcal{D}_{\hat{p}_k \hat{x}_j}[\hat{\rho}_{\rm S}]
		\right)
= 
\sum_{jk} \mathcal{S}_{jk}
		\left(
		i\mathcal{D}_{\hat{b}_j^\dagger\hat{b}_k^\dagger}[\hat{\rho}_{\rm S}(t)] -
		i\mathcal{D}_{\hat{b}_j\hat{b}_k}[\hat{\rho}_{\rm S}(t)] + \rm{h.c.}
		\right)
\end{align}
Numerical calculations showed that for our chosen parameters $\mathcal{S}_{jk} \ll \Gamma_{jk}\bar{n} \ll \Omega_1$ for the parameters in \cref{tab:parameters}. Thus, the above term proportional to $\mathcal{S}_{jk}$ can be neglected by a rotating wave approximation. 
\par
Finally, we comment on the thermal damping term in
\cref{equ:2-particle-Born-Markov-Master-Equation}, $\mathcal{D}_{\rm
th}[\hat{\rho}]$. To model the thermal damping and decoherence due to other sources than the
wire, one can in principle include in \cref{equ:linearizedHamiltonian-grouped} a Caldeira-Leggett bath of harmonic oscillators that is independent of the wire degrees of freedom ($\hat{\mathbf{f}}(\mathbf{r}, \omega)$), where the bath evolution and its interaction with the system are described by
\begin{align}
	\hat{H}^{\rm (CL)}_{\rm B}
	& =
	\sum_{\lambda}
	\left[
	\frac{\hat{P}_{\lambda}^2}{2m_{\lambda}}
	+
	\frac{1}{2}m_{\lambda}\omega_{\lambda}^2
	\hat{X}_{\lambda}^2
	\right]
	\\
	\hat{H}^{\rm (CL)}_{\rm int}
	& =
	\sum_{j=1}^2 \hat{X}_j
	\sum_{\lambda} c_{j\lambda}\hat{X}_{\lambda}.
\end{align} 
Here, $c_{j\lambda}$ is the coupling strength, and $\hat{P}_{\lambda}$ and $\hat{X}_{\lambda}$ denote the momentum and position operators of bath mode $\lambda$ respectively.
We can then apply the same procedure as above, which results only in the addition of $\mathcal{D}_{\rm
th}[\hat{\rho}]$ in \cref{equ:2-particle-Born-Markov-Master-Equation} as long as the validity assumptions made in the main text hold
\cite{kolodynskiAddingDynamicalGenerators2018}.

\section{Master equation rates}

\begin{figure}
	\includegraphics[width=\textwidth]{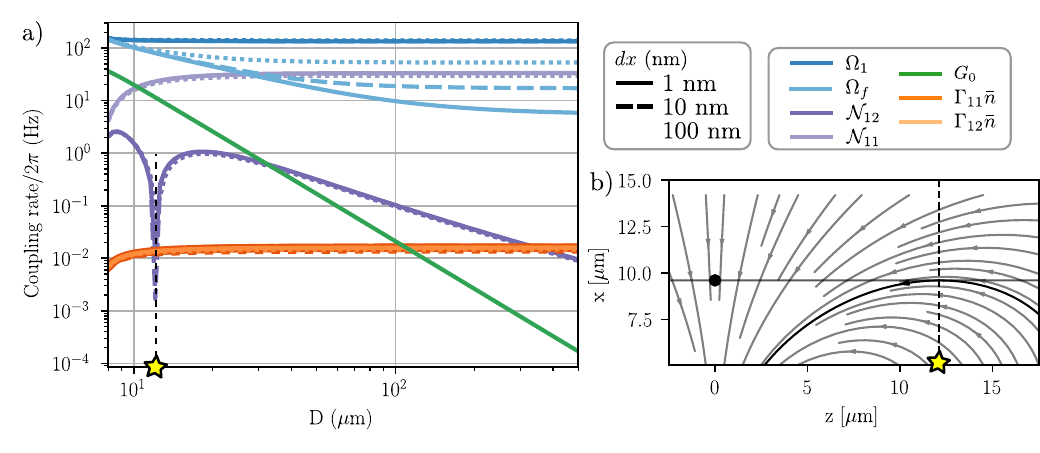}
	\caption{(a) All the rates of the master equation as a function of interparticle separation $D$ and distance to instability $dx$. (b) the wire-mediated coupling $\mathcal{N}_{12}$ is given by the x-components of the field along the horizontal line, which explains the zero-crossing in (a) (see text for details)}
	\label{fig:Rates}
\end{figure}

For completeness, we give a more detailed look at all the master equation rates in \cref{fig:Rates}(a). We show the renormalized frequencies $\Omega_1$ and $\Omega_f$, the local and non-local wire-mediated coupling rates $\mathcal{N}_{11}$ and $\mathcal{N}_{12}$ respectively, as well as both wire-induced decoherence rates $\Gamma_{12}$ and $\Gamma_{11}$ as a function of distance. Here, it becomes visible that the main contributor to the $dx$ dependence of $G_{\rm tot}$ in \cref{fig:Fig2} is the change in $\Omega_f$. Additonally, the decoherence rates $\Gamma_{11}$ and $\Gamma_{12}$ remain roughly equal, with $|\Gamma_{11} - \Gamma_{12}|/\Gamma_{11} < 2\times10^{-7}$, and do not decay with distance in this range. The wire-induced coupling $\mathcal{N}_{12}$ changes sign at around 12.1 $\mu$m in \cref{fig:Rates}(a). To see why this happens, note that the Green's tensor components $\partial_{X_j} \partial_{X_k} g^{\rm M}(\mathbf{R}^{\rm equ}_{j}, \mathbf{R}^{\rm equ}_{k}, \Omega_1)$ appearing in the definition of $\mathcal{N}_{12}$ \cref{eq:ratesN12} can be interpreted as the x-component of the wire-mediated electric field at $\mathbf{R}^{\rm equ}_{j}$ of a point dipole at $\mathbf{R}^{\rm equ}_k$ that is pointing into the x-direction. In \cref{fig:Rates}(b), we show this vector field in the XZ-plane. The zero-crossing happens due to this field pointing perpendicular to the x-axis.\par

\section{Asymptotic Expansion of Green's tensor}
\label{sec:asymptoticExpansion}

To understand the scaling behaviour that is seen for the wire-mediated coupling rate in \cref{fig:Fig2}, we aim to asymptotically expand the Green's function appearing in \cref{eq:ratesN12} for $D\gg d$. We make two simplifying assumptions: (i) we assume the perfect conductor limit $\Omega_1 \ll \sigma/\epsilon_0$, which implies $Q_m(k d/2, \Omega_1) \approx 1$, and (ii) we take the geometry of \cref{sec:rates}, i.e. $x=x'$, $y=y'=0$ and $z-z' = D$. We define the integrand of the Fourier integral in \cref{eq:Greensfunction}
\begin{align}
	f_m(k,x) =
	\frac{I_m(kd/2)}{K_m(kd/2)}k^2\times
	\begin{cases}
		\frac{1}{2}K_1(kx)^2                    & \text{for $m=0$}	 \\
		 (K_{m+1}(kx) - m K_m(kx)/(kx))^2 & \text{for $m>0$}
	\end{cases}
\end{align}
so that we can write
\begin{align}\label{eq:greensfunction-appendix}
	\partial_x \partial_{x'}
	g^{\rm M}(\mathbf{R}, \mathbf{R}',\omega)\vert_{x=x'} =
	-\frac{1}{\pi^2 \epsilon_0}
	\sum\limits_{m=0}^\infty
	\int_0^\infty dk
	\cos(kD)f_m(k,x).
\end{align}
The function $f_m(k,x)$ has no poles for any $m\geq 0$, $k>0$ and is exponentially damped as $k \gg 1/d$ as long as $x>d/2$ (which is physically always the case). As we will see below, the main behaviour of \cref{eq:greensfunction-appendix} then comes from the behaviour of $f_m(k,x)$ when $k D \ll 1$. The integrand in that limit can be approximated \cite{NIST:DLMF} as
\begin{align}
	f_m(k,x)
	\approx
	\begin{cases}
		-1/(2 \ln(k d/2)x^2)                           & \text{ for }m=0 \\
		\left(\frac{d}{2x}\right)^{2m} \frac{m}{2 x^2} & \text{ for }m>0
	\end{cases}.
\end{align}
Since the $m>0$ contributions are constant to zeroth order in $k$, a partial integration shows that these terms will decay as $1/D^2$. Thus, we focus on the $m=0$ term in \cref{eq:greensfunction-appendix}.\par
We use a modified Watson's lemma which is proved in Ref.\cite{kupinVersionWatsonLemma2021}. To be able to apply it, we first split $\cos(kD) = \frac{1}{2}(e^{ikD} + e^{-ikD})$, then substitute $k$ with $\pm i k'$. This transforms the integration axis to the positive and negative imaginary axis respectively. We can rotate it back to the positive real axis by noting that (i) the integral enclosing the upper/lower quadrants is zero since $K_0(z)$ has no zeros in these regions (and therefore $f_0(k)$ no poles) and (ii) the integrand vanishes on the large quarter circles. Therefore, we can exchange the integral over the positive/negative imaginary axes with the same along the positive real axis. All this yields
\begin{align}
	\partial_x \partial_{x'}
	g^{\rm M}(\mathbf{R}, \mathbf{R}',\omega)\vert_{x=x'}
	\approx
	-\frac{i}{2\pi^2\epsilon_0}
	\int_0^\infty dk\, e^{-kD}
	\left[
		f_0(ik,x)-f_0(-ik,x)
	\right].
\end{align}
The integrand satisfies the assumptions in Ref.\cite{kupinVersionWatsonLemma2021} and thus it reduces to 
\begin{align}
	\partial_x \partial_{x'}
	g^{\rm M}(\mathbf{R}, \mathbf{R}',\omega)\vert_{x=x'}
	\approx
	\frac{1}{4\pi \epsilon_0 x^2 D \ (\ln^2(2D/d) + \pi^2/4)}
	\propto \frac{1}{D \ln^2(D)},
\end{align}
where the last proportionality is valid for $D\gg d$.

\section{Input-output relations}\label{sec:input-output}

In this section we derive the coupling rate between the particles and and the measured input-output modes. We start off with the extended system of \cref{sec:measurementsetup}, which includes (i) the two particles, (ii) the wire degrees of freedom and their interaction with the particles, and (iii) two cavity modes, each coupled optomechanically to one of the particles, as well as to the continuum of free traveling modes. Following the input-output formulation of Ref.~\cite{alessioserafiniQuantumContinuousVariables2023} we write the total Hamiltonian in the frame rotating at the cavity frequency as 
\begin{align}\label{equ:Hamiltonianwithcavity+travelingmodes}
	\hat{H}_{\rm tot}/\hbar = \hat{H}/\hbar + \sum_{j=1}^2 g \hat{x}_j \hat{\mathcal{X}}_j + \sum_{j=1}^2 i\int_\mathbb{R} ds \sqrt{\kappa} \delta(s-t) (\hat{a}_j^\dagger \hat{b}_{s,j} - \hat{a}_j \hat{b}_{s,j}^\dagger),
\end{align}
where $\hat{H}$ is the particle-wire Hamiltonian in \cref{equ:linearized-H}. The traveling modes are described
 through ladder operators $\hat{b}_{s,j}$, where $j=1,2$ is the index labeling the cavity they couple to, and where $s$ is a time index. These operators are
assumed to satisfy the white noise commutators $\comm{\hat{b}_{s,j}}{\hat{b}^\dagger_{r,k}} = \delta_{jk} \delta(s-r)$. The time-dependent interaction Hamiltonian can be interpreted as follows: A traveling mode is injected into the cavity at time $t$ and interacts with the system for a time $\tau$ that is smaller than all relevant system time-scales $1/\tau \gg g, \Omega_1, G_{\rm tot}$ and can thus be assumed to be a single instant of time. The mode is then scattered through an output port. The temporal label of the operators $\hat b_{s,j}$ marks the time $s$ when the interaction happens. \par
The Heisenberg equations of motion for the cavity- and traveling modes read
\begin{align}\label{equ:Heisenberg-cavity-travelmode}
	\dot{\hat{b}}_{s,j} & = -\sqrt{\kappa} \delta(s-t) \hat{a}_j(t)    ,      \\
	\dot{\hat{a}}_j     & = -i g \hat{x}_j - \sqrt{\kappa} \hat{b}_{t,j}(t).\label{equ:EOM-cavity}
\end{align}
Integrating the first equation yields
\begin{align}\label{equ:input-output1a}
 \hat{b}_{s,j}(t) = \hat{b}_{s,j}(0) - \sqrt{\kappa}\,\Theta(t - s)\,\hat{a}_j.
\end{align} 
Inserting this result into \cref{equ:EOM-cavity} with the half-maximum convention $\Theta(0) = 1/2$ gives
\begin{align}
	\label{equ:input-output1b}
	\dot{\hat{a}}_j(t) = -i g \hat{x}_j -\frac{\kappa}{2} \hat{a}_j(t) + \sqrt{\kappa}\hat{b}_{t,j}(0).
\end{align}
Now, we define
\begin{align}
	\hat{b}_{t,j}(t') =
	\begin{cases}
		\hat{b}^{\rm in}_{j}(t)   & \text{, if $t' < t$ (before interaction)} \\
		-\hat{b}^{\rm out}_{j}(t) & \text{, if $t'> t$ (after interaction)}
	\end{cases}.
\end{align}
The $\pi$-phase convention in front of $\hat{b}^{\rm out}_{j}$ signals that the input and output modes have opposite propagation direction. With that \cref{equ:input-output1a,equ:input-output1b} become what are known as the input-output relations in the literature,
\begin{align}
	 & \hat{b}^{\rm in}_{j} + \hat{b}^{\rm out}_{j} = \sqrt{\kappa} \hat{a}_j,\label{equ:input-output2a}                              \\
	 & \dot{\hat{a}}_j = -i g \hat{x}_j -\frac{\kappa}{2} \hat{a}_j + \sqrt{\kappa}\hat{b}^{\rm in}_{j}. \label{equ:input-output2b}
\end{align}
In the bad cavity regime ($\kappa \gg g$), the cavity will approximately stay in its steady state so that $\dot{\hat{a}}_j \approx 0$, resulting in 
\begin{align}
	\hat{a}_j \approx -i \frac{2 g}{\kappa} \hat{x}_j + \frac{2}{\sqrt{\kappa}}\hat{b}^{\rm in}_{j}. \label{equ:input-output-bad-cavity}
\end{align}
This is equivalent to \cref{equ:input-output-bad-cavity-quadrature-a,equ:input-output-bad-cavity-quadrature-b} in the main text. Furthermore, using \cref{equ:input-output-bad-cavity} to replace the cavity modes in the Hamiltonian \cref{equ:Hamiltonianwithcavity+travelingmodes}, evaluating the Dirac delta function $\delta(s-t)$ explicitly, and inserting \cref{equ:input-output1a} at $x=t$, one arrives directly at \cref{eq:measurement-Hamiltonian} in the main text.

\section{Derivation of the Riccati equation}\label{sec:ricatti-derivation}

This section is essentially a special case and collection of the more general derivation of the Kalman filter equations for continuously measured Gaussian states in Ref.~\cite{alessioserafiniQuantumContinuousVariables2023}. This formulation is chosen both because of minimal use of stochastic calculus as well as providing an intuitive explanation why the quantum filtering equation for Gaussian dynamics reduces to the classical Kalman filter equations. In contrast to Ref.~
\cite{alessioserafiniQuantumContinuousVariables2023}, we focus on the special cases of homodyne measurement of the phase quadrature $\mathcal{Y}^{\rm out}_\pm$ instead of a generaldyne measurement of a general auxiliary mode quadrature. The idea is to first derive an infinitesimal update of the Gaussian state describing both the particles and the output modes. After the infinitesimal evolution there are non-zero covariances between these two subsystems. A projective measurement on the output modes will then result in an updated state for the particles. As we will show, the evolution of the continuously updated particle state is described by the classical Kalman filter equation.\par
Let us proceed by deriving the mean and covariances of particles and output modes after an infinitesimal time-evolution. The full master equation including the input coupling of the previous section (here shown in its form before the normal mode transformation) reads
\begin{align}
	\begin{split}
		\frac{d\hat{\rho}}{dt} = \mathcal{L}[\hat{\rho}] =
		-\frac{i}{\hbar}\Big[
			\hat{H}_{\rm eff}
			+\sum_{j=1}^2 \frac{4g}{\sqrt{\kappa}} \hat{\mathcal{X}}^{\rm in}_j
			\hat{x}_j\, ,
			\hat{\rho}
			\Big]
		+ \mathcal{D}_{\rm M}[\hat{\rho}]
		+ \mathcal{D}_{\rm th}[\hat{\rho}].
	\end{split}\label{equ:Master-Equation-Input-Coupling}
\end{align}
The key relation that is needed to calculate the infinitesimal update for the first and second moments is the infinitesimal commutator for the input modes 
\begin{align}\label{equ:infinitesimal-commutator}
	[\hat{\mathcal{X}}^{\rm in}_j, \hat{\mathcal{Y}}^{\rm in}_k]dt = 2i \delta_{jk}.
\end{align}
It follows directly from the white-noise commutator of the input fields $[\hat{b}^{\rm in}_{j}(t),\hat{b}^{\rm in\dagger}_{k}(t')] = \delta_{jk}\delta(t-t')$ which in term of the quadratures reads $[\hat{\mathcal{X}}^{\rm in}_{j}(t),\hat{Y}^{\rm in}_{k}(t')] = 2 i \delta_{jk}\delta(t-t')$. The time-integrated input quadratures then obey
\begin{align}
	\left[
		\int_t^{t+dt} dt'\, \hat{\mathcal{X}}^{\rm in}_j(t'),
		\int_t^{t+dt} dt''\, \hat{\mathcal{Y}}^{\rm in}_k(t'')
	\right]
	= 2i\delta_{jk}dt
\end{align}
which -- as long as $dt \ll \tau$ (where $\tau$ is the interaction time between input mode and cavity from the previous section) -- reduces to \cref{equ:infinitesimal-commutator}. \cref{equ:infinitesimal-commutator} suggests that we should rewrite the equations in terms of the quantum Wiener increments $\hat{x}^{\rm in}_j = \hat{\mathcal{X}}^{\rm in}_j \sqrt{dt}$ and $\hat{y}^{\rm in}_j = \hat{\mathcal{Y}}^{\rm in}_j \sqrt{dt}$ which fulfill canonical commutation relations.
Using cyclicity of the trace, the commutators of $\hat{x}_j$, $\hat{p}_j$, and the Wiener increments above, one finds the equations of motion using
\begin{align}\label{equ:operator-calculation}
	\langle \hat{\mathcal{O}}\rangle (t+dt) = \Tr\left(
		\hat{\mathcal{O}}  
		\left(
			\mathds{1} + dt \mathcal{L}[\hat{\rho}] + \frac{dt^2}{2} \mathcal{L}^2[ \hat{\rho}]
		\right)
	\right),
\end{align}
where we have to go to second order because essentially $dt^2 \hat{\mathcal{X}}^{\rm in}_{j} \hat{\mathcal{X}}^{\rm in}_{k} = dt \hat{x}^{\rm in}_{j}\hat{x}^{\rm in}_{k}$.
We write the updated covariance and mean vectors into block form, splitting system S -- the measured mechanical modes -- and bath B -- the input modes. That is, we write the total phase space vector as  $\hat{\bm{\eta}}_{\rm tot} = (\hat{x}_1, \hat{p}_1, \hat{x}_2, \hat{p}_2, \hat{x}^{\rm in}_1, \hat{y}^{\rm in}_1, \hat{x}^{\rm in}_2, \hat{y}^{\rm in}_2)^T$. Similarly, we can define the total covariance matrix of motion + input modes as $\bm{\Sigma}_{\rm tot} = \frac{1}{2}\langle\{ \hat{\bm\eta}_{\rm tot}, \hat{\bm\eta}_{\rm tot}\}\rangle$, which can be split into blocks as
\begin{align}
	\bm{\Sigma}_{\rm tot}(t) =
	\begin{pmatrix}
		\bm{\Sigma}_S(t)    & \bm{\Sigma}_{SB}(t) \\
		\bm{\Sigma}_{SB}(t) & \bm{\Sigma}_B(t)
	\end{pmatrix}.
\end{align}
Here $\bm\Sigma_S(t)$ and $\bm\Sigma_B(t)$ are the covariance matrices of the two particles and of the two input modes, respectively, whereas $\bm\Sigma_{SB}(t)$ represents their correlations.
We assume no correlations between the motion of the particles and the input mode at the start time $t$ of the infinitesimal
time-evolution, i.e. $\bm\Sigma_{SB}(t) = 0$. We also use our assumption that the input mode is in a vacuum state, i.e.  $\bm{\Sigma}_{B}(t) = \mathds{1}$. Next, the normal mode transformation given by \cref{equ:input-output-bad-cavity-quadrature-a,equ:input-output-bad-cavity-quadrature-b,equ:normal-mode-trafo-a,equ:normal-mode-trafo-b} is performed. If the normal modes are initially uncorrelated, i.e. $\langle \{\hat{\bm\eta}_+(t), \hat{\bm\eta}_-(t) \}\rangle=0$, they will continue to be, so that the evolution equations for the two normal mode covariances decouple into
\begin{align}\label{eq:infinitesimal-sigma-evolution}
	\bm{\Sigma}^\pm(t+dt) = (\bm{\Sigma}_S^\pm(t) \oplus \bm{\Sigma}_{B}^\pm)(t) + \sqrt{dt} \begin{pmatrix}
		                                                                                   0                   & \bm{\Sigma}_{C}^\pm(t) \\
		                                                                                   \bm{\Sigma}_{C}^\pm(t) & 0
	                                                                                   \end{pmatrix}
	- \delta\bm{\Sigma}_S^\pm(t) \oplus \delta\bm{\Sigma}_{B}^\pm(t) dt + O(dt^{3/2}),
\end{align}
with
\begin{align}
	 & \bm{\Sigma}_{C}^\pm(t)= - \frac{4 g_\pm}{\sqrt{\kappa}} 
	 \begin{pmatrix}
		0 & \langle \hat{x}_\pm^2\rangle(t)
		\\
		1 & \langle \hat{x}_\pm \hat{p}_\pm \rangle(t) + \langle \hat{p}_\pm \hat{x}_\pm \rangle(t)
	\end{pmatrix}, \\
	 & \delta\bm{\Sigma}_{\rm B}^\pm=
	\mathrm{diag}\left(
	0, \frac{64 g_\pm^2}{\kappa} \langle \hat{x}_\pm^2 \rangle(t)
	\right)                                                                                                                                       \\
	 & \delta\bm{\Sigma}_S^\pm(t) = A_\pm \bm{\Sigma}_S^\pm(t) + \bm{\Sigma}_S^\pm(t) A_\pm + (2\bar{n}+1) V_\pm + H_\pm
\end{align}
where we have used the definitions of $A_\pm$, $V_\pm$, $g_\pm$ and $H_\pm$ from \cref{eq:Riccati,equ:EOMSA,equ:EOMSB}. Importantly, one can see from \cref{eq:infinitesimal-sigma-evolution} that we are left with a non-zero correlation between input-mode and system after the interaction as well as the addition $H_\pm$ to the usual diffusion term in $\delta\bm{\Sigma}_S$. \par
In view of \cref{equ:input-output-bad-cavity-quadrature-b}, the position information of the particles is encoded in the output phase quadratures $\hat{\mathcal{Y}}^{\rm out}_\pm$. The goal is now to determine the system state after a projective measurement on this output quadrature yielding a result $y_0$ (this can e.g. be realised approximately by homodyne measurements). Note that after the infinitesimal time-evolution in \cref{eq:infinitesimal-sigma-evolution} input quadratures should be replaced by output quadratures. Since the normal modes decoupled in \cref{eq:infinitesimal-sigma-evolution}, we will omit the corresponding $\pm$-indices and consider one system mode S coupled to one output mode B in the following calculation.\par
Let us determine the system state after a measurement of the output mode $\hat{\mathcal{Y}}^{\rm out}$ has yielded the result $y_0$. We make use of the Fourier-Weyl representation of our two-mode Gaussian state after the infinitesimal time-evolution. From now on, we omit all time-dependences since no time-evolution is occuring in the following argument. The state is characterized by its covariance matrix $\bm{\Sigma}$ and mean $\bar{\bm{\eta}}=(\bar{x}_S,\bar{y}_S, \bar{x}_B,\bar{y}_B)^T = \langle\hat{\bm{\eta}}\rangle$, where $\hat{\bm\eta} = (\hat x_S, \hat p_S, \hat x_B, \hat y_B)^T$. 
\begin{equation}\label{equ:Gaussian-Weyl-representation}
	\hat\rho = \frac{1}{(2\pi)^2} \int_{\mathbb{R}^{4}}d\bm{\eta}
	e^{-\frac{1}{4}\bm{\eta}^T\Lambda^T\bm{\Sigma}\Lambda \bm{\eta} + i\bm{\eta}^T\Lambda^T\bar{\bm{\eta}}} \hat{D}_{\bm{\eta}}.
\end{equation}
This is a Gaussian integral over displacement operators $\hat{D}_{\bm{\eta}} = e^{i \bm{\eta}^T \Lambda \hat{\bm{\eta}}}$ where $\Lambda$ denotes the 4x4 symplectic form defined by $\comm{\eta_j}{\eta_k} = i \Lambda_{jk}$. After the measurement the displacement operator partially collapses to  
\begin{equation}
	\bra{y_0} \hat{D}_{\bm{\eta}} \ket{y_0}_B =
	e^{-2ix_B y_0} \delta(y_B) \hat{D}_{\bm{\eta}_S},
\end{equation}
Thus, the system state is projected to
\begin{align}\label{equ:Born-update-1}
	\bra{y_0}\hat{\rho} \ket{y_0} = \frac{1}{(2\pi)^{2}}\int_{\mathbb{R}^{4}} d\bm{\tilde{\eta}}\, e^{
			-\frac{1}{4}\left(
			\tilde{\bm{\eta}}_S^T\bm{\Sigma}_S \tilde{\bm{\eta}}_S
			+ 2 \tilde{\bm{\eta}}_B^T\bm{\Sigma_{SB}}^T\tilde{\bm{\eta}}_S
			+ \tilde{\bm{\eta}}_B^T\bm{\Sigma}_B \tilde{\bm{\eta}}_B
	\right) }
	e^{
			i \tilde{\bm{\eta}}_S^T\bar{\bm{\eta}}_S
			+ i \tilde{\bm{\eta}}_B \bar{\bm{\eta}}_B}
	e^{- 2 i x_B^T y_0 }
	\hat{D}_{\bm{\eta}_S} \delta(y_B),
\end{align}
where we have defined \(\tilde{\bm{\eta}} = \Lambda \bm{\eta}\). To simplify the expression, we evaluate the Dirac-delta and collect all exponentials with exponents proportional to $\tilde{\bm{\eta}}_B$, which is equal to $(0, -2x_B)^T$ if $y_B=0$. This leaves us with 
\begin{align}
	\bra{y_0}\hat{\rho} \ket{y_0} = \frac{1}{(2 \pi)^{2}} \int_{\mathbb{R}^{2}}d\tilde{\bm{\eta}}_S
	e^{
			-\frac{1}{4}
			\tilde{\bm{\eta}}_S^T\bm{\Sigma}_S \tilde{\bm{\eta}}_S}
	e^{
			i \tilde{\bm{\eta}}_S^T\bar{\bm{\eta}}_S}
	\hat{D}_{\bm{\eta}_S} 
	\int_{\mathbb{R}} dx_B\,
	e^{
			-\left(
			- x_B  (\bm{\Sigma_{SB}}^T\tilde{\bm{\eta}}_S)^y  \,
			+ x_B\langle \hat y_B^2 \rangle\, x_B
			\right) }
	e^{- 2 i x_B \bar{y}_B}
	e^{- 2 i x_B y_0 },
\end{align}
where we denote with $\bm{\eta}^y$ the phase-components of a phase-space vector, e.g. $\bm{\eta}_S^y = y_S$. The integral over $x_B$ is simply a Gaussian integral, inserting the known solution yields
\begin{align*}
	\bra{y_0}\hat{\rho} \ket{y_0} =
	\frac{1}{(2 \pi)} \int_{\mathbb{R}^{2}}
	d\tilde{\bm{\eta}}_S
	e^{
			-\frac{1}{4}
			\tilde{\bm{\eta}}_S^T\bm{\Sigma}_S \tilde{\bm{\eta}}_S}
	e^{
			i \tilde{\bm{\eta}}_S^T\bar{\bm{\eta}}_S}
	\hat{D}_{\bm{\eta}_S} 
	\frac{1}{\sqrt{\langle \hat{y}_k^2 \rangle}}
	e^{\frac{1}{4}
			\left(
			(\bm{\Sigma}_{SB}^T\tilde{\bm{\eta}}_S)^y
			- 2i (\bar{y}_B - y_0)
			\right)^2
			/(2 \langle \hat y_B^2 \rangle)}.
\end{align*}
This is still a Gaussian state which can be seen explicitly by collecting the terms according to their powers of $\tilde{\bm{\eta}}_S$. The exponent can then be written as
\begin{align}
	&-\frac{1}{4}
	\tilde{\bm{\eta}}_S^T
	\left(
	\bm{\Sigma}_S
	- \bm{\Sigma}_{SB} \Pi_y\, \bm{\Sigma}_B^{-1} \Pi_y\, \bm{\Sigma}_{SB}^T
	\right)
	\tilde{\bm{\eta}}_S
	+ i \tilde{\bm{\eta}}_S^T
	\left(
	\bar{\bm{\eta}}_S
	- \bm{\Sigma}_{SB} \Pi_y\, \bm{\Sigma}_B^{-1} \Pi_y
	(\bar{\bm{\eta}}_B - \bm{\eta}_0)
	\right)
	+\text{const.},
\end{align}
where we have defined the projector onto the momentum subspace $\Pi_y = \mathrm{diag}(0,1,0,1)$ and defined 
$\bm{\eta}_0 = (0, y_0)^T$.
From the above expression we can read the updated system covariance matrix after the measurement, 
\begin{equation}\label{eq:sigma-conditional-update}
	\bm{\Sigma}_S \mapsto
	\bm{\Sigma}_S
	- \bm{\Sigma}_{SB} \Pi_y\, \bm{\Sigma}_B^{-1} \Pi_y\, \bm{\Sigma}_{SB}^T.
\end{equation}
 Additionally, the system average gets shifted as well,
\begin{equation}\label{eq:sigma-mean-update}
	\bar{\bm{\eta}}_S \mapsto
	\bar{\bm{\eta}}_S
	- \bm{\Sigma}_{SB} \Pi_y
	\bm{\Sigma}_B^{-1} \Pi_y (\bar{\bm{\eta}}_B - \bm{\eta}_0).
\end{equation}
Here, we can also observe what is classically known as the Kalman filter matrix 
$K = \bm{\Sigma}_{SB} \Pi_y \bm{\Sigma}_B^{-1} \Pi_y$. 
Combining the infinitesimal evolution of the covariance matrix under coupling to the system, \cref{eq:infinitesimal-sigma-evolution}, with the update after every measurement in \cref{eq:sigma-conditional-update}, one obtains the classical Kalman filter equation \cref{eq:Riccati}.
The fundamental reason why the classical equations emerge is that the Born rule in \cref{equ:Born-update-1} essentially implements a Bayesian update and the Gaussian state is analogous to a classical probability distribution on phase space.
Then, for Gaussian prior probability distributions and Markovian dynamics, it is known that the Bayesian filter reduces to a Kalman filter \cite{bechhoeferControlTheoryPhysicists2021}.

\end{document}